\pdfoutput=1
\documentclass[12pt,twoside,journal,draftclsnofoot,onecolumn]{IEEEtran}

\usepackage{amsmath,amsfonts,bm,amssymb,psfrag,ifthen,color,subfigure}
\usepackage{mathrsfs}
\usepackage[justification=centering,font=footnotesize]{caption}
\usepackage{todonotes}

\allowdisplaybreaks[4]

\usepackage[dvips]{epsfig}
\usepackage[dvips]{graphicx}
\usepackage{amsmath,amsfonts,bm,amssymb,psfrag,ifthen,color,subfigure}
\usepackage{algpseudocode}
\usepackage{algorithm}
\usepackage{algorithmicx}
\usepackage{ifthen}
\usepackage{setspace}
\usepackage{cite}
\usepackage{url}
\usepackage{longtable}
\usepackage{stfloats} 
\usepackage{graphics,booktabs,color,subfigure}
\usepackage{float}
\usepackage{diagbox}

\usepackage[dvips]{epsfig}
\usepackage[dvips]{graphicx}
\usepackage{amsmath,amsfonts,bm,amssymb,psfrag,ifthen,color,subfigure,amsthm}
\usepackage{algpseudocode}
\usepackage{algorithm}
\usepackage{algorithmicx}
\usepackage{setspace}
\usepackage{cite}
\usepackage{url}
\usepackage{longtable}
\usepackage{stfloats} 
\usepackage{graphics,booktabs,color,subfigure}
\usepackage{float}
\usepackage{diagbox}
\usepackage{multirow}

\allowdisplaybreaks[4]

\newcommand{\tabincell}[2]{\begin{tabular}{@{}#1@{}}#2\end{tabular}}

\newtheorem{Lemma}{Lemma}
\newtheorem{Proposition}{Proposition}
	\usepackage{graphicx}

	\usepackage{graphicx}
	\usepackage{todonotes} 
	
\begin{document}
		\title{Task-oriented Explainable Semantic Communications}
		
		\author{Shuai Ma, Weining Qiao, Youlong Wu, Hang Li, Guangming Shi,~\IEEEmembership{Fellow,~IEEE}, Dahua Gao, Yuanming Shi, Shiyin Li, and Naofal Al-Dhahir,~\IEEEmembership{Fellow,~IEEE}
			
			\thanks{Shuai Ma is with
				Pengcheng Laboratory, Shenzhen, 518066, China (e-mail: mash01@pcl.ac.cn).}

		}
		
		\maketitle
		\begin{abstract}

			Semantic communications utilize the transceiver computing resources
			to alleviate scarce transmission resources, such as bandwidth and energy.
			Although the conventional deep learning (DL) based designs may achieve certain transmission
			efficiency, the uninterpretability issue of extracted features
			is the major challenge in the development of semantic communications.
			In this paper, we propose an explainable and robust semantic communication framework
			by incorporating the well-established bit-level communication system,
			which not only extracts and disentangles
			features into independent and semantically interpretable features,
			but also only selects task-relevant features for transmission,
			instead of all extracted features.
			Based on this framework, we derive the optimal input for
			rate-distortion-perception theory, and derive both lower and upper bounds on the semantic channel capacity.
			Furthermore, based on the
			$\beta $-variational autoencoder ($\beta $-VAE), we propose a practical explainable
			semantic communication system design, which
			simultaneously achieves semantic features selection and is robust against
			semantic channel noise.
			We further design a real-time wireless mobile semantic
			communication proof-of-concept prototype. Our simulations and experiments demonstrate that our proposed
			explainable semantic communications system can significantly improve transmission
			efficiency, and also verify the effectiveness of our proposed robust semantic transmission scheme.

		\end{abstract}
		
		\IEEEpeerreviewmaketitle
		\section{Introduction}
		

		With the advent of augmented reality (AR), virtual reality (VR),
		holographic communications, autonomous vehicular networks, and industrial
		Internet of Things (IIoT),
		it is envisioned that existing networks may soon reach
		a resource bottleneck due to stringent requirements \cite{Lueth_2020,Antoniou_2021}, such as ultra-high data rate, ultra-reliability, and low latency. To meet the above-mentioned requirements, investigations on the sixth generation communications (6G) are well underway and promise more powerful capacities than the fifth-generation communications (5G) \cite{Saad_INW_2020}.
		From the first generation communications (1G) to 5G, the communication networks primarily focus on finding new resources and technologies to expand the channel capacity\cite{Strinati_TVT_2019}. One approach is to seek the usage of large bandwidth, such as terahertz (THz) communications and visible light communication (VLC). Another approach is to explore the spatial domain, like ultra-massive MIMO and intelligent metasurfaces.
		However, given the hardware and physical limitations, the channel capacity may not keep increasing at the rate we desire to satisfy the aforementioned beyond-5G applications \cite{Bao_ICST_2022,Niu_arXiv_2022}.

		In recent years, semantic communications, in which
		only task-relevant information is extracted and transmitted to the receiver,
		have received
		increasing attention by both academia and the industry		 \cite{Zhang_Engineering_2022,Kountouris_CM_2021,Sana_CCNC_2022,Shi_CM_2021,Luo_WC_2022,Bao_INSW_2011,Yener_TCCN_2018}.
		Rather than increasing the channel capacity as in the conventional techniques, semantic
		communications exploit the computing power at the transceivers to alleviate the cost of transmission resources. The classic Shannon information theory focuses on
		``How accurately can
		the symbols be transmitted?'', which ignores the
		meaning of the transmitted messages.
		Instead,
		 semantic
		communications \cite{Weaver_1949} consider ``How precisely do the transmitted symbols convey
		the desired meaning?''
		Thus, it is possible to improve the system efficiency at the semantic level, not only at the pure bit level.

		The classical separation theorem \cite{Shannon_1948} states that,
		as
		the data size goes to infinity, separating source coding and channel coding can achieve the optimal performance over a memoryless communication channel.
		However, for finite number of bits transmission, the performance of such separated structure will degrade.
		This issue also arises in semantic communications. Various deep learning (DL) based
		joint source-channel coding (JSCC) schemes have been investigated for
		text \cite{Bao_INSW_2011,Farsad_ICASSP_2018,Xie_TSP_2021}, image \cite{Hu_arXiv_2022,Bourtsoulatze_TCCN_2019,Kurika_JSAC_2020,Yang_TCCN_2022,Choi_PMLR_2019,Xu_TCSVT_2021}, speech
		\cite{Tong_GLOBECOM_2021,Weng_JSAC_2021}, and
		multimodal data \cite{Xie_WCL_2021}
		transmission.		
		Specifically, for text semantic transmission,
		the JSCC schemes have been designed by exploiting architectures like the recurrent neural network
		(RNN) \cite{Farsad_ICASSP_2018}, Transformer \cite{Xie_TSP_2021,Zhou_arxiv_2022}, autoencoder (AE)\cite{Jiang_arXiv_2021}, adaptive Universal Transformer\cite{Zhou_WCL_2021}, and deep neural network (DNN) \cite{Xie_JSAC_2021}.
		For image semantic transmission,
		a masked auto-encoder (MAE) architecture with Transformer was designed in \cite{Hu_arXiv_2022} to combat adversarial samples noise.
		Convolutional neural networks (CNNs) based JSCC schemes were designed for
		the time-invariant and fading wireless channels in \cite{Bourtsoulatze_TCCN_2019}.
		Neural error correcting
		and source trimming (NECST) codes were studied in \cite{Choi_PMLR_2019}.
		For finite bit transmission, an
		attention
		DL based JSCC method was designed in \cite{Xu_TCSVT_2021}.
		By exploring the channel output feedback,
		an AE-based JSCC scheme was developed in \cite{Kurika_JSAC_2020} to improve the quality of image transmission.
		By combining an AE with
		orthogonal frequency division multiplexing (OFDM), a
		JSCC wireless image transmission scheme was presented in \cite{Yang_TCCN_2022} over multipath fading channels.
		By leveraging reinforcement
		learning (RL), a
		joint semantics-noise coding (JSNC) mechanism was designed in \cite{Lu_arxiv_2021}.
		A DNN based
		JSCC scheme was designed in \cite{Yang_arXiv_2021} for adaptive rate control in wireless image transmission.
		Based on AE,
		a SNR-adaptive deep JSCC scheme is
		proposed in \cite{Ding_ICASSP_2021} for multi-user wireless image transmission.
		To tackle the variational information bottleneck, the authors in \cite{Shao_JSAC_2022} investigated task-oriented communication
		for edge inference, where a low-end edge device extracts the feature vector of a local data sample and transmits to a powerful
		edge server for processing.
		Besides, for the speech semantic transmission,
		AE based wave-to-vector architecture and squeeze-and-excitation (SE) attention network have been developed in \cite{Tong_GLOBECOM_2021} and \cite{Weng_JSAC_2021}, respectively.
		For visual
		question answering, the memory-attention-composition neural network was designed in \cite{Xie_WCL_2021} for multi-modal data
		semantic communications.

  However, most of the existing works on semantic communications \cite{Farsad_ICASSP_2018,Xie_TSP_2021,Hu_arXiv_2022,Bourtsoulatze_TCCN_2019,Kurika_JSAC_2020,Yang_TCCN_2022,Choi_PMLR_2019,Xu_TCSVT_2021,Tong_GLOBECOM_2021,Weng_JSAC_2021,Xie_WCL_2021}
		are based on DL techniques, in which the DL model is basically a black box.
 Thus, the extracted semantic feature vectors in these works are unexplainable (hidden) representations, and the uninterpretability of the extracted features restricts further processing and exploitation of semantic features. For example, due to the uninterpretability,
the unintended features will also be transmitted to the receiver, which wastes transmission resources
and reduces the efficiency of semantic communications.

	Moreover, most of the existing semantic communication investigations \cite{Farsad_ICASSP_2018,Xie_TSP_2021,Hu_arXiv_2022,Bourtsoulatze_TCCN_2019,Kurika_JSAC_2020,Yang_TCCN_2022,Choi_PMLR_2019,Xu_TCSVT_2021,Tong_GLOBECOM_2021,Weng_JSAC_2021,Xie_WCL_2021}
 completely redesign the source and channel module over the conventional system, which are impractical and not compatible with the existing communication networks. Because there is a large number of practical standards and hardware for 5G physical layer, it will lead to a huge waste of resources and costs by replacing  physical layer techniques with DL-based semantic JSCC techniques.
 Therefore, how to design efficient and 5G-compatible practical semantic communications is a critical issue.




		To address the above the two key challenges of the semantic communications, we propose an explainable and robust semantic communication framework in this paper, which is compatible with existing communication systems. We show that the proposed framework can achieve a higher transmission efficiency than the existing inexplicable semantic communication systems.
		The main contributions of this paper are summarized as follows:
		\begin{itemize}

			\item We propose an explainable and easy-to-implement semantic communication framework	 based on the bit-level communication systems,
			which includes a novel semantic encoder, as well as the corresponding decoder,
			feature selection and semantic channel. The innovation of the proposed
			framework is threefold: i) The semantic encoder/decoder aims to, not only
			extract the independent and explainable semantic information as \emph{semantic source coding},
			but also alleviate the ambiguity of the semantic information influenced
			by the quantization and channel noise as \emph{semantic channel coding};
			ii) The feature selection module follows the semantic encoder,
			to choose only the task-relevant features for transmission,
			which can further reduce the transmission load; iii) The framework has an explicit definition of semantic channels, which incorporates the key modules of the bit-level communication systems. Specifically, the semantic channel takes both quantization error (or noise) and physical channel noise into account since those noise sources may lead to semantic information ambiguity, and the semantic channel capsulizes the conventional bit-level communication systems, which implies that the proposed framework can be more easily implemented compared to the JSCC schemes.

			\item Then, we   propose two information-theoretic metrics for our semantic communication framework. In terms of the    information compression of the semantic encoder,
			we derive the  the optimal distribution
of the reconstruction signal of the  rate-distortion-perception function for
			semantic information extraction.
		  Moreover, to quantify the semantic information transmission, we derive both upper and lower bounds  for the semantic channel capacity, which are shown to be tight when the quantization noise tends to zero.

			\item Based on our framework, we further propose a feasible design of the explainable semantic communication system. Specifically, this design includes a robust $\beta$-VAE lightweight unsupervised learning network, where a weighted parameter $\beta$ is added to the Kullback-Leibler (KL) divergence term of the variational autoencoder (VAE) network loss function, in order to make the latent representations effectively disentangled. Moreover, to
			enhance transmission robustness, the semantic channel noise is added to the extracted features during semantic networks training.

			\item Finally, we implement the above semantic communication design, and propose a wireless mobile semantic communication proof-of-concept prototype.
			Applying the
			portable
			Raspberry Pi 4 Model B and Wi-Fi, the developed prototype can run the proposed robust $\beta$-VAE semantic system in real time.
			Our experiments demonstrate that our proposed semantic communication system can achieve better performance than existing benchmarks.

		\end{itemize}

		The rest of this paper is organized as follows. The
		explainable semantic communications framework is presented in Section II.
		Section III provides the information-theoretic metrics of semantic communications.
		In Section IV, we propose a $\beta$-VAE based robust and explainable semantic communications system.
		In Section V, we present the semantic
		communication system prototype design and implementation.
		In Section VI, we evaluate the proposed explainable semantic
		communication system. Finally, we conclude the paper in Section VII. Table I and II presents the
means of the key notations and key acronyms in this paper, respectively.

		\begin{table}[H]
	\caption{Key Notations and Meanings}
	\label{tablepar}
	\centering
	\begin{tabular}{|c|l|}
		\hline
		\rule{0pt}{8pt}Variables & Meanings \\ \hline
		\rule{0pt}{7.5pt}$S = \left\{ {{s_k}} \right\}_{k = 1}^K$ & \tabincell{c}{Semantic information with $K$ features
} \\ \hline
		\rule{0pt}{7.5pt}${s_k}$ & \tabincell{c}{The $k$th semantic feature} \\ \hline
		\rule{0pt}{7.5pt}$X $ & \tabincell{c}{Source data}\\ \hline
		\rule{0pt}{7.5pt}$ Z = \left\{ {{z_l}} \right\}_{l = 1}^L$ & \tabincell{c}{The extracted semantic feature vector with $L$ features }\\ \hline
		\rule{0pt}{7.5pt}$z_l$ & \tabincell{c}{The $i$-th extracted semantic feature}\\ \hline
		\rule{0pt}{7.5pt}${\mathcal{L}} $ & \tabincell{c}{Semantic feature index set}\\ \hline
		\rule{0pt}{7.5pt}${\mathcal{L}_{{\rm{sel}}}}$ & \tabincell{c}{Selected semantic feature index set}\\ \hline
		\rule{0pt}{7.5pt}${X_{\rm{s}}} = {\left\{ {{z_l}} \right\}_{l \in {\mathcal{L} _{\rm{sel}}}}} $ & \tabincell{c}{Selected semantic features}\\ \hline
		\rule{0pt}{7.5pt}${Y_{\rm{s}}} = \{{\widehat z}_l\}_{l \in {{\cal L}_{{\rm{sel}}}}}$ & \tabincell{c}{Estimated semantic features }\\ \hline
	\rule{0pt}{7.5pt}$\widehat {{Z}}$ & \tabincell{c}{Reconstructed feature set}\\ \hline
	\rule{0pt}{7.5pt}$\widehat {{X}}$ & \tabincell{c}{Decoded data }\\ \hline
	\end{tabular}
\end{table}

		\begin{table}[H]
	\caption{Key Acronyms and Meanings}
	\label{tablepar}
	\centering
	\begin{tabular}{|c|l|}
		\hline
		\rule{0pt}{8pt}Acronyms & Meanings \\ \hline
		\rule{0pt}{7.5pt}JSCC & \tabincell{c}{Joint source-channel coding } \\ \hline
		\rule{0pt}{7.5pt}VAE & \tabincell{c}{Variational autoencoder} \\ \hline
		\rule{0pt}{7.5pt}KL & \tabincell{c}{Kullback-
Leibler}\\ \hline
		\rule{0pt}{7.5pt}ANGC & \tabincell{c}{Additive non-Gaussian noise channel}\\ \hline
		\rule{0pt}{7.5pt}ELBO & \tabincell{c}{Evidence lower bound}\\ \hline
		\rule{0pt}{7.5pt}GPU & \tabincell{c}{Graphics processing unit}\\ \hline
		\rule{0pt}{7.5pt}PSNR & \tabincell{c}{Peak signal-to-noise ratio}\\ \hline
	\end{tabular}
\end{table}

		\section{Explainable Semantic Communication Framework}

		Most existing studies replace the traditional source coding and channel coding modules by deep learning-based studies source-channel coding, which greatly changes the structure of the existing communication systems. In this paper, we propose a semantic communication framework incorporating the key modules of the conventional communication system (e.g., 5G).
				\begin{figure*}[htbp]
			\centering
			\includegraphics[width=0.8\textwidth]{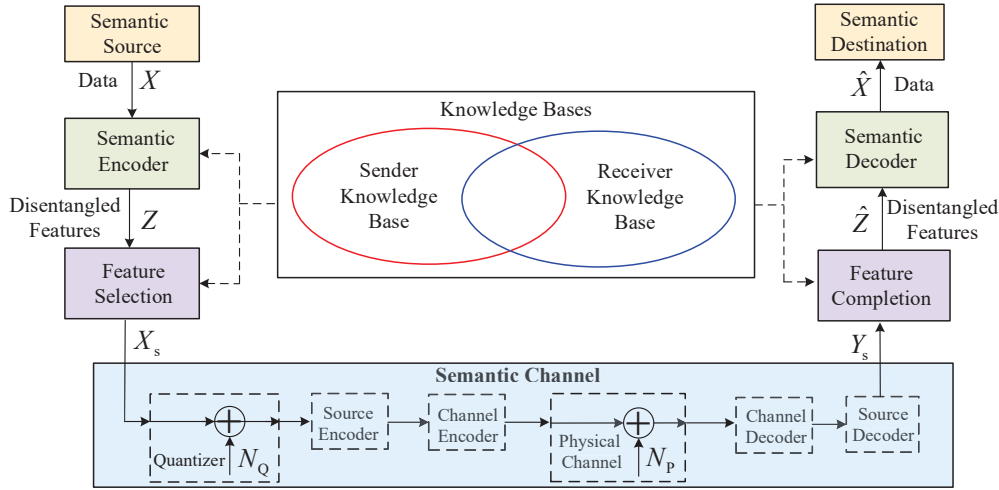}
			\caption{Explainable and robust semantic communications framework }
			\label{PTP_SM}
		\end{figure*}

		As shown in Fig.~1, the proposed explainable semantic communication framework includes a semantic source, sender knowledge base, semantic encoder, semantic channel, receiver knowledge base, semantic decoder, and semantic destination.
		Note that the proposed framework introduces a semantic-level transmission on the top of bit-level transmission.
		Clearly, such a framework does not require the extra redesign over the existing physical standards, protocols and products, which makes the application of semantic communications more practical.
		Next, we will describe each module in detail.

		\subsection{Knowledge Bases}
		The knowledge base contains all the necessary information that can facilitate the communication at the semantic level. Specifically, the	
knowledge base includes background knowledge and training
dataset. The background knowledge is used to facilitate the semantic feature
extraction and selection in the semantic transmitter. The training dataset is used for training the parameters of the semantic
encoder and decoder.
		The sender may choose different semantic knowledge bases according to different tasks, scenarios and recipients. For example, when the communication is triggered between people in different countries, it may be necessary to sample multiple language databases.
		In general, the sender and the receiver share some common knowledge, which
		may act as a
		special kind of side information to improve coding efficiency.

%

		%

		\subsection{Semantic Sources} The semantic source produces original data, such as pictures, videos, voices, and texts. The generated data contains certain semantic information to be shared with the semantic destination.
		Assume that the semantic information includes $K$ features $S = \left\{ {{s_k}} \right\}_{k = 1}^K \sim {p_{{\rm{sou}}}}\left( s \right)$ (data generative factors),
		where ${s_k}$ denotes the $k$th semantic feature, and the joint probability distribution is ${p_{{\rm{sou}}}}\left( {{s}} \right)$. Further, assume that the $K$ features $\left\{ {{{s}_k}} \right\}_{k = 1}^K$ are independent, i.e., ${p_{{\rm{sou}}}}\left( {{s}} \right) = \prod\limits_{k = 1}^K {{p_{{\rm{sou}}}}\left( {{s_k}} \right)}$, where ${{p_{{\rm{sou}}}}\left( {{s_k}} \right)}$ denotes the probability distribution of ${{s_k}}$.
		Thus, the entropy of the semantic source is given as
		\begin{align}\label{H_S}
			{{H}}\left( {S} \right) = - \sum\limits_{k = 1}^K {{p_{{\rm{sou}}}}\left( {{s_k}} \right){{\log }_2}{p_{{\rm{sou}}}}\left( {{s_k}} \right)}.
		\end{align}

		The semantic source can generate the data $X \sim {p_{{\rm{data}}}}\left( x \right)$, which can be images, text, sound or video.
		Generally, the generated data need to include both the intended features and some redundant features
		to make the whole semantic data complete.
		Thus, the data generation is defined as ${p_{{\rm{s2d}}}}\left( {{{x}}|\left\{ {{{s}_k}} \right\}_{k = 1}^K} \right)$, and the probability distribution function (PDF) of data $X$ is given as
		\begin{align}\label{H_X}
			{p_{{\rm{data}}}}\left( x \right) = \sum\limits_{{s_1},...,{s_K}} {{p_{{\rm{s2d}}}}\left( {x|\left\{ {{s_k}} \right\}_{k = 1}^K} \right)\prod\limits_{k = 1}^K {{p_{{\rm{sou}}}}\left( {{s_k}} \right)} } .
		\end{align}
		The entropy of the semantic data $x$ is given as
		\begin{align}
			{{H}}\left( X \right) = - \sum\limits_x {{p_{{\rm{data}}}}\left( x \right){{\log }_2}{p_{{\rm{data}}}}\left( x \right)} .
		\end{align}
		
		Based on \eqref{H_S} and \eqref{H_X}, $H\left( {{X}} \right)$ can be further expressed as
		\begin{align}H\left(X\right) = H\left( S \right) + H\left( {X|S} \right) - H\left( {S|X} \right).
		\end{align}

		\subsection{Semantic Encoder}
		Based on the knowledge base, the generated message $X$ will be processed by the semantic encoder, which is a joint semantic source and channel encoder. More specifically, it
		extracts semantic information or the
		semantic features of the message $X$, and outputs the disentangled and explainable features $Z$, which can be viewed as a semantic source encoder. On the other hand, in order to reduce the ambiguity incurred by the quantization error and channel noise, the semantic encoder needs to improve the robustness against the semantic channel noise, which can be viewed as a semantic channel encoder.

		%

		The semantic encoder extracts a low-dimensional semantic features vector $Z \sim {p_{{\rm{fea}}}}\left( z \right)$ from the data ${X}$.
		Let ${p_{{\rm{d2f}}}}\left( {z|x} \right)$ denote the conditional PDF of the feature $z$ given data $x$. Thus, the PDF of the extracted feature (sub-vectors) ${p_{{\rm{fea}}}}\left( z \right)$ is given as
		\begin{align}{p_{{\rm{fea}}}}\left( z \right) = \sum\limits_x {{p_{{\rm{d2f}}}}\left( {z|x} \right){p_{{\rm{data}}}}\left( x \right)}.\end{align}
		
		The encoder is required to regulate the extracted features into $L$ independent features $Z = \left\{ {{z_l}} \right\}_{l = 1}^L$, which satisfy
		\begin{align}{p_{{\rm{fea}}}}\left( {{z}} \right) = \prod\limits_{l = 1}^L {{p_{{\rm{fea}}}}\left( {{z_l}} \right)},
		\end{align}
		where $p\left( {{{z}_{l}}} \right)$ denote the PDF for feature ${{z}_l}$.
		In summary, the extracted feature ${\bf{z}}$ is required to have $L$ disentangled interpretable semantic features $\left\{ {{{{z}}_l}} \right\}_{l = 1}^L$,
		whose corresponding neural network output is explainable and understandable by the human.
		For convenience, we let ${\mathcal{L}} \buildrel \Delta \over = \left\{ {1,...,L} \right\}$ denote the index set of the disentangled semantic features. Note that such a requirement can be met if the semantic encoder is designed in a sophisticated manner. In Section IV, we will introduce a feasible system design that has such capability.

		\subsection{Feature
			Selection}

		It should be noted that the obtained features $Z$ could contain more information than what the receiver is interested in.
		Thus, after extracting the disentangled features $\{z_l\}_{l=1}^L$, only the subset of features $\{z_l\}_{l=1}^L$ that are of interest to the receiver should be transmitted, and the rest of the features can be viewed as the ``redundancy". We will present more discussions of this issue via experiments in Section VI.

		Given the task requirement, let
		${\mathcal{L}_{{\rm{sel}}}} \subseteq {\mathcal{L}}$ denote the selected feature index set, then the selected set of features is given as
		\begin{align}{X_{\rm{s}}} = {\left\{ {{z_l}} \right\}_{l \in {\mathcal{L} _{\rm{sel}}}}}.
		\end{align}

		Thus, feature selection will reduce the amount of data sent,
		and the corresponding reduction is ${\left\{ {{z_l}} \right\}_{l \in {\cal L}\backslash {{\cal L}_{{\rm{sel}}}}}}$. Then, ${X_{\rm{s}}}$ will be sent to the semantic channel.

		\subsection{Semantic Channel}

		After the feature selection module, the task-oriented features are selected and ready to send. Since the quantization error and channel noise both could incur semantic information ambiguity, we define the semantic channel with channel law $p({y_s|x_s})$ as a virtual channel including the signal quantizer and the bit-level communication system, as shown in Fig. 1. Here, ${Y_{\rm{s}}} = \{{\widehat z}_l\}_{l \in {{\cal L}_{{\rm{sel}}}}}\sim {p_{{\rm{r}}}}\left( {{y_{\rm{s}}}} \right)$ represents the set of estimated features after the transmissions over the semantic channel, and ${\widehat z}_l$ denotes the estimated feature of $z_l$.

 Generally, the semantic noise could include various factors including source errors, feature extraction errors, knowledge base ambiguities, adversarial injections, quantization noise, physical channel noise, etc. In our framework, the semantic channel noise $N_s$ is the distortion between
 the selected semantic feature ${X_{\rm{s}}} = {\left\{ {{z_l}} \right\}_{l \in {\mathcal{L} _{\rm{sel}}}}}$ and the estimated semantic feature ${Y_{\rm{s}}} = \{{\widehat z}_l\}_{l \in {{\cal L}_{{\rm{sel}}}}}$, which mainly depends on the quantization noise and physical channel noise.

		\subsubsection{Quantization Noise}
		The quantization noise is
		caused by the traditional communication operation modules, such as the source encoder (or decoder) or channel encoder (or decoder), which may also lead to semantic ambiguity.
		In order to reduce the number of transmitted bits, the semantic feature ${{{\bf{x}}_{\rm{s}}}}$ will be converted to a compressible binary stream
		using few bits.
		To represent ${{{\bf{x}}_{\rm{s}}}}$ with a finite number of bits, we need to map it to a discrete space. Specifically, a finite quantizer
		maps the semantic feature ${{{x}}_{\rm{s}}}$ to ${{{x}}_{\rm{b}}}$, whose
		values are then quantized to $M$ levels $C = \left\{ {{c_1},...,{c_M}} \right\}$, i.e., \begin{align}{{{x}}_{\rm{b}}} = {\rm{Quan}}\left( {{{{x}}_{\rm{s}}}} \right),\end{align}
		where ${\rm{Quan}}\left( \cdot \right)$ is a quantization operator.
		Since the number of dimensions $\dim \left( {{{{x}}_{\rm{b}}}} \right)$ and the number of
		levels $L$ are finite, the entropy of quantized semantic data is given as
		\begin{align}H\left( {{{{x}}_{\rm{b}}}} \right) \le \dim \left( {{{{x}}_{\rm{b}}}} \right){\log _2}M.
		\end{align}
		In this paper, we consider the uniform distributed quantization noise ${{{n}}_{\rm{Q}}}$, i.e.,
		\begin{align}{p_{{{\rm{N}}_{\rm{Q}}}}}\left( x \right) = \frac{1}{{b - a}},a \le x \le b,
		\end{align}
		where $a$ and $b$ are the lower and upper bounds of quantization noise ${{{N}}_{\rm{Q}}}$.

		\subsubsection{Physical Channel Noise}
		
		The physical channel noise exists ubiquitously
		in physical communications and is caused by
		physical channel impairments, such as additive
		white Gaussian noise (AWGN), interference, etc. It is noted
		that the errors caused by channel propagation
		usually occur before channel decoding and can
		be corrected by channel decoding.
		Assume that the physical channel noise ${{N}}_{\rm{P}}$ follows a Gaussian distribution with zero-mean and variance $\sigma _{\rm{P}}^2$, i.e.,
		\begin{align}{p_{{{\rm{N}}_{\rm{P}}}}}\left( x \right) = \frac{1}{{{\sigma _{\rm{P}}}\sqrt {2\pi } }}\exp \left( { - \frac{{{x^2}}}{{2\sigma _{\rm{P}}^2}}} \right).
		\end{align}

		\subsection{Feature Completion}
 	After obtaining the estimated features ${{{Y}}_{\rm{s}}}={\left\{ {{{\widehat z}_l}} \right\}_{l \in {{\cal L}_{{\rm{sel}}}}}}$ through the semantic channel transmission, the destination will use the estimated features and side information in the knowledge base, to compute the target function of the task. Although the unintended features subset ${\left\{ {{z_l}} \right\}_{l \in \mathcal{L}\backslash {\mathcal{L}_{{\rm{sel}}}}}}$ are not transmitted, the receiver
		may generate the corresponding unintended features ${\left\{ {{{\widehat {z}}_{l}}} \right\}_{l \in \mathcal{L}\backslash {\mathcal{L}_{{\rm{sel}}}}}}$ by exploiting the knowledge base.		
		Then, by
		combining intended features ${\left\{ {{{\widehat {z}}_{l}}} \right\}_{l \in {\mathcal{L}_{{\rm{sel}}}}}}$ and unintended features ${\left\{ {{{\widehat {z}}_{l}}} \right\}_{l \in \mathcal{L}\backslash {\mathcal{L}_{{\rm{sel}}}}}}$, we may obtain the completed semantic features $\widehat Z = {\left\{ {{{\hat z}_l}} \right\}_{l \in \mathcal{L}}}$ with distribution ${{p_{{\rm{rfea}}}}\left( {\widehat z} \right)}$.
 For example,
considering a semantic communication system for staff clothing image transmission, the intended semantic features of the receiver are clothing features, and the receiver
 is not interested in the staff's gender, skin color, and hairstyle. Therefore, the receiver can generate unintended semantic features based on the shared knowledge base, such as the staff's gender, skin color and hairstyle. Note that, the generated unintended semantic features at the receiver may be different from the corresponding features of the image at the transmitter.
 Then, the receiver combines the received clothing features with its own generated unintended features.


		\subsection{Semantic Decoder}
		The semantic decoder aims to recover the data from the disentangled features $\widehat Z$ that are semantic explainable, which is the inverse function of the semantic encoding. Again, this inverse function needs the help of the knowledge base for model training such that the
		decoder can ``understand" the features $\widehat Z$.

		Similar to the encoding process, we use conditional PDF ${p_{{\rm{f2d}}}}\left( {\widehat x|\widehat {{z}}} \right)$ to describe the semantic decoding process. The PDF of the decoded data is ${p_{{\rm{rdata}}}}\left( {\widehat x} \right)$, and the decoded data is ${\widehat {{x}}}$.
		The data reconstruction for a given feature vector is given as
		\begin{align}{p_{{\rm{rdata}}}}\left( {\hat x} \right) = \sum\limits_{\widehat z} {{p_{{\rm{f2d}}}}\left( {\hat x|\widehat z} \right){p_{{\rm{rfea}}}}\left( {\widehat z} \right)}. \end{align}

		\subsection{Semantic Destination} Finally, the receiver recovers the semantic information based on the decoded data ${\widehat {{X}}}$, and the corresponding process can be described by ${p_{{\rm{d2s}}}}\left( {\widehat {{s}}|\widehat x} \right)$, where the final semantic information is denoted by ${\widehat {s}}$. At last, the probability of such semantic information can be written as
		\begin{align}{p_{{\rm{des}}}}\left( {\widehat s} \right) = {p_{{\rm{d2s}}}}\left( {\widehat s|\widehat x} \right){p_{{\rm{rdata}}}}\left( {\widehat x} \right).\end{align}

		So far, we have presented the complete semantic communication framework. The key modules are the semantic encoder and the feature selection. Their functions can be realized by the careful model design. We will present a detailed system design in Section IV, which is a feasible realization of this framework.

		\section{Information-Theoretic Metrics of Semantic Communications }
		
		In this section, we propose two metrics for the framework illustrated by Fig. 1. Here, we focus on two procedures: the encoding and the transmission.

			\subsection{Rate-Distortion-Perception Function}

		The semantic encoding may include many different tasks, and these tasks may have relevant or different criteria. For example, there is data distortion for the traditional data reconstruction task, and distribution distortion for generative learning tasks.
		
	 Let $p(x)$ be the distribution of the input source, $r(\widehat x)$ be the distribution of the reconstruction signal, and $q(\widehat x|x)$ be a conditional distribution on $\mathcal{X}\times {\mathcal{X}} $.
			The information rate-distortion-perception function $R(D,P)$ \cite{Blau_ICML_2019} for a source $X\sim p(x)$ is defined as
			\begin{subequations}\label{R_D_P}
				\begin{align}
					R(D,P) &= \mathop {\min }\limits_{q( {\widehat x|x} )} I ( {X;\widehat X} )\\
					{\rm{s.t.}}&\quad {\mathbb{E}}\left[ {\Delta \left( {x,\widehat x} \right)} \right] \le D,\label{R_D_P_b1}\\
					&\quad{d}\left( {p( x ),r( {\widehat x} )} \right) \le P,\label{R_D_P_c1}\\
					&\quad\sum\limits_{\widehat x} {q( {\widehat x|x} )} = 1,\forall x\in\mathcal{X}.\label{R_D_P_d1}
				\end{align}
			\end{subequations}
			where the distortion function $\Delta: \mathcal{X}\times { \mathcal{X}}\to \mathbf{R}^+ $ satisfying $\Delta(x,\widehat x) = 0$ if $x=\widehat x$, and perception function $d\big(p(x),r(\widehat x)\big)$ is a non-negative divergence between probability distributions $p(x)$ and $r(\widehat x)$ satisfying $d(p,q)= 0$ if $p(x)=r(x)$.

		

		So far, for a general source, the optimal distribution of the reconstruction signal $r(\widehat x)$
of problem \eqref{R_D_P} has not been derived yet.
 For a binary source, the three-way
tradeoff between rate, distortion, and perception was investigated in \cite{Blau_ICML_2019} with Hamming distance distortion and total-variation distance perception. While
for a Gaussian source, the achievable distortion-perception region was established
 in \cite{Zhang_NIPS_2021} under squared error distortion and squared Wasserstein-2 distance.

	Hence, we investigate how to find the optimal of $R ( {D,P})$ for
 a general source under the mean square distortion (i.e., $\Delta(x,\widehat x)=|x-\widehat x|^2 $)
 and KL divergence perception (i.e., $\quad{d}\left( {p\left( x \right),r( {\widehat x}) } \right)=\quad{d}_\text{KL}\left( {p\left( x \right),r( {\widehat x}) } \right)\triangleq\sum_{x} p\left( x \right)\log \frac{p(x)}{r({x})}$).
 We first introduce the following lemma.
%
%
%
%

		\begin{Lemma}\label{BAoptimalDistr}
			Consider the mean square distortion (i.e., $\Delta(x,\widehat x)=|x-\widehat x|^2 $) and KL divergence perception (i.e., $\quad{d}\left( {p\left( x \right),r( {\widehat x}) } \right)=\sum_{x} p\left( x \right)\log \frac{p(x)}{r({x})}$).
			The corresponding optimal distribution $q^*(\widehat x|x)$ to problem \eqref{R_D_P} for a given output distribution $r(\widehat x)>0$ is
			\begin{IEEEeqnarray}{rCl}\label{Qoptimal}
				q^*\left( {\widehat x|x} \right) = \frac{{r( {\widehat x}) }}{{\widetilde \gamma \left( x \right)}}\exp \left( {\mu \frac{{p\left( x \right)}}{{r( {\widehat x}) }} - \alpha {{\left( {x - \widehat x} \right)}^2}} \right),
			\end{IEEEeqnarray}
			where $\widetilde \gamma \left( x \right) = \sum\limits_{\widehat x} {r( {\widehat x}) \exp \left( {\mu \frac{{p\left( x \right)}}{{r( {\widehat x}) }} - \alpha {{\left( {x - \widehat x} \right)}^2}} \right)} $.
			The corresponding optimal distribution $r^*(\widehat x)$ to \eqref{R_D_P} for a given conditional distribution $q(\widehat x|x)>0$ is
			\begin{IEEEeqnarray}{rCl}\label{Roptimal}
				r^*(\widehat x) = \sum_{{x}}p(x)q(\widehat x|x).
			\end{IEEEeqnarray}
		\end{Lemma}
		\emph{Proof:} Please find the proof in Appendix A.\qed
		
		Using Lemma \ref{BAoptimalDistr}, we can apply a process of alternating minimization, called the Blahut–Arimoto algorithm \cite{Cover_Book}. Specifically, in the initialization setup, choose some positive values $\alpha, \mu$ and the initial output distribution $ {r}^{(0)}(\widehat x)$. In each iteration $k$, compute the optimal $q^{(k)}(\widehat x|x)$ according to \eqref{Qoptimal} for given $r^{(k-1)}(x)$, and then compute the optimal $r^{(k)}(x)$ according to \eqref{Roptimal}.

		\subsection{Lower and Upper Bounds on Semantic Channel Capacity}
		
		The channel capacity quantifies the maximum rate of information transmission for the considered system. According to the framework in Fig. 1, we define the semantic channel capacity as the maximum semantic information that can be transferred through the semantic channel $p({y_{\rm{s}}|x_{\rm{s}}})$. Following the standard achievability and converse proof techniques, we obtain the semantic channel capacity in our framework as:
		\begin{IEEEeqnarray}{rCl}
		{C_{\rm{s}}}= \max_{p(x_{\rm{s}})} I(X_{\rm{s}};Y_{\rm{s}}).
\end{IEEEeqnarray}
		
		In the conventional bit-level wireless communication system, the channel capacity is usually represented by the Shannon capacity formula with additive Gaussian distributed noise.
 In our framework, the semantic channel noise $N_s$ mainly depends on the quantization noise and physical channel noise, and follows non-Gaussian distribution in general.
 Thus, in our framework, the semantic channel is an additive non-Gaussian noise channel (ANGC), and we assume that
 the estimated semantic features ${Y_{\rm{s}}}$ can be represented as
\begin{align}
{Y_{\rm{s}}}
 &= {{\rm{X}}_{\rm{s}}} + {N_s}.
	\end{align}
	 Although the specific distribution of ${N_s}$ is unknown,
 the variance of the semantic noise ${n_{\rm{s}}}$ can be obtained by measurement.
 In this paper, we assume that the covariance of the semantic noise ${n_{\rm{s}}}$ is $\sigma _{\rm{s}}^2$.



 Due to the non-Gaussian distributed semantic noise ${n_{\rm{s}}}$, the classic Shannon capacity formula (based on Gaussian distributed noise) cannot be directly applied to the semantic channel.
		To derive the semantic channel capacity, we first define equivalent Gaussian distributed semantic channel noise ${\overline {{N}} _{\rm{s}}} \sim \mathcal{N}\left( {0,\sigma _{\rm{s}}^2}\right)$ with the same variance as ${{{N}}_{\rm{s}}}$.
		Then, based on the equivalent semantic channel noise ${\overline {{N}} _{\rm{s}}}$, the received signal of semantic channel ${\overline {{Y}} _{\rm{s}}}$ is given as
		\begin{align}{\overline {{Y}} _{\rm{s}}} = {{{X}}_{\rm{s}}} + {\overline {{N}} _{\rm{s}}}.
		\end{align}
		Therefore, the channel capacity of the equivalent semantic channel is given as
		\begin{align}
			{C_{{\rm{s}},{\rm{eq}}}} = \frac{1}{2}\log \left( {1 + \frac{{{P_{{{\rm{x}}_{\rm{s}}}}}}}{{\sigma _{\rm{s}}^2}}} \right),
		\end{align}
		where ${P_{{{\rm{x}}_{\rm{s}}}}}$ denote the power of transmitted semantic data ${X_{\rm{s}}}$.

		\begin{Proposition}[Lower and upper bounds on the semantic channel capacity]
			With the non-Gaussian distributed channel noise, the semantic channel capacity ${C_{\rm{s}}}$is bounded by \cite{Ihara}
			\begin{align}{C_{{\rm{s,eq}}}} \le {C_{\rm{s}}} \le {C_{{\rm{s,eq}}}} + {d_{{\rm{KL}}}}\left( {{p_{{{\rm{n}}_s}}}\left( x \right),{p_{{{\overline {\rm{n}} }_{\rm{s}}}}}\left( x \right)} \right),\end{align}
			where ${d_{{\rm{KL}}}}\left( {{p_{{{\rm{n}}_s}}}\left( x \right),{p_{{{\overline {\rm{n}} }_{\rm{s}}}}}\left( x \right)} \right) = \int_{ - \infty }^{ \infty } {{p_{{{\rm{n}}_{\rm{s}}}}}\left( x \right)\log \frac{{{p_{{{\rm{n}}_{\rm{s}}}}}\left( x \right)}}{{{p_{{{\overline {\rm{n}} }_{\rm{s}}}}}\left( x \right)}}} {\rm{d}}x$.
		\end{Proposition}
		\begin{figure}[htbp]
			\centering
			\begin{minipage}[t]{0.45\textwidth}
				\centering
				\includegraphics[width=\textwidth]{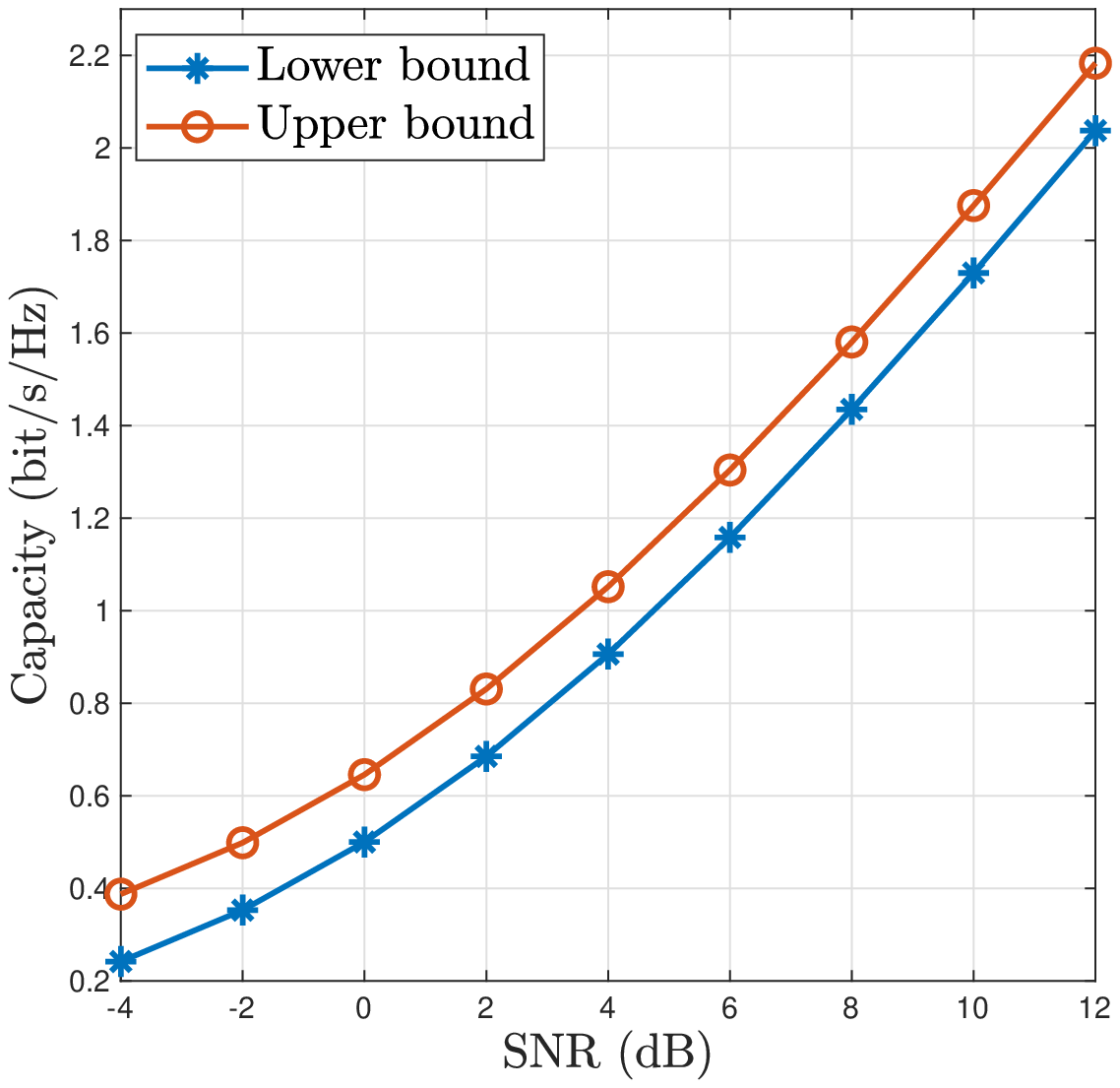}
				\vskip-0.2cm\centering {\footnotesize (a)}
			\end{minipage}
			\begin{minipage}[t]{0.45\textwidth}
				\centering
				\includegraphics[width=\textwidth]{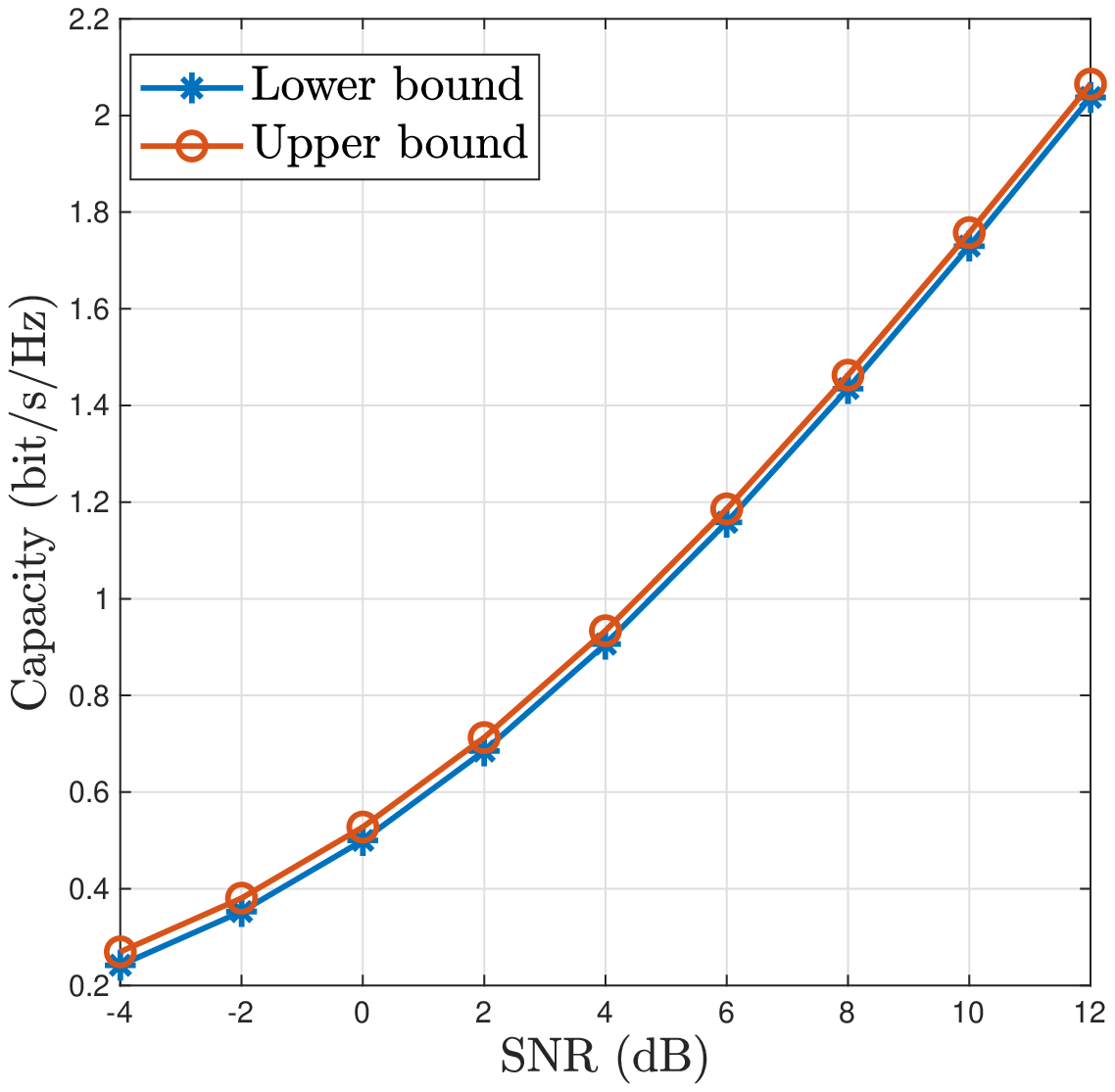}
				\vskip-0.2cm\centering {\footnotesize (b)}
			\end{minipage}
			\caption{
				(a) Lower and upper bounds of semantic channel capacity versus SNR with $a=-1$, $b=1$ and $\sigma _{\rm{P}}^2 = 0.01 $;
				(b) Lower and upper bounds of semantic channel capacity versus SNR with $a=-0.3$, $b=0.3$ and $\sigma _{\rm{P}}^2 = 0.01$.}
			\label{capactiy of rate}
		\end{figure}

		At last, we illustrate our theoretical results on the semantic channel capacity via numerical simulation.
		Fig. \ref{capactiy of rate} (a) and (b) show the lower bound and the upper bound in (22) on semantic channel capacity versus SNR with semantic noise parameters $a=-1$, $b=1$ and $\sigma _{\rm{P}}^2 = 0.01 $, and semantic noise parameters $a=-0.3$, $b=0.3$ and $\sigma _{\rm{P}}^2 = 0.01$, respectively.
 	Fig. \ref{capactiy of rate} (b) shows that the gap between the upper bound and lower bound is
 less than that in Fig. \ref{capactiy of rate} (a). The reason is that when the
		 the KL divergence between semantic noise ${\bar{N}_s}$ and the equivalent
		semantic channel noise ${{{\overline {\rm{n}} }_{\rm{s}}}}$ tends to $0$, i.e., ${d_{{\rm{KL}}}}\left( {{p_{{{\rm{n}}_s}}}\left( x \right),{p_{{{\overline {\rm{n}} }_{\rm{s}}}}}\left( x \right)} \right) \to 0$, the gap between the lower bound and the upper bound in (22) tends to 0.

		\section{ $\beta$-VAE based Robust and Explainable Semantic Communication System}

\begin{figure}
			\centering
			\includegraphics[width=1.0\textwidth]{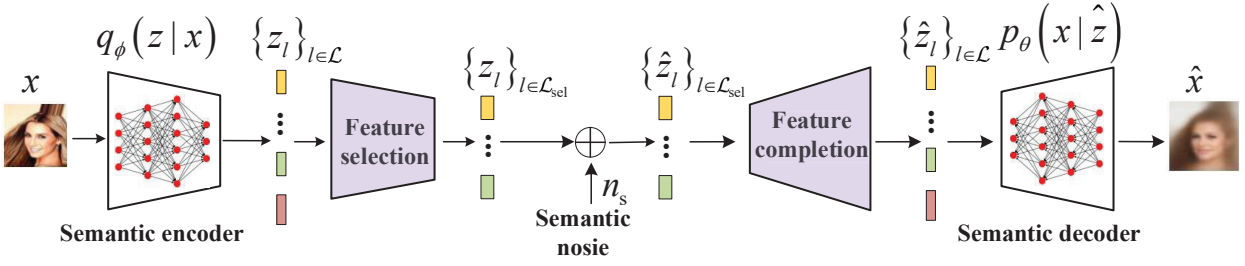}
			\caption{Proposed $\beta$-VAE based explainable semantic communication system}
			\label{beta_SM}
		\end{figure}

		In this section, we present a feasible and efficient system design based on the proposed framework given in Fig. 1. Here, we propose a robust $\beta$-VAE based semantic communications system, as shown in Fig. \ref{beta_SM}, which disentangles the hidden
		representation vector into multiple independent and semantically interpretable
		of features.

		\subsection{ Robust $\beta$-VAE based Semantic Encoder/Decoder}
		By exploiting a generative VAE model \cite{Higgins_ICLR_2017}, we first optimize
		the semantic encoder ${{q_\phi }\left( {{{z}}|{{x}}} \right)}$ with parameter set $\phi$,
		and the semantic decoder ${p_\theta }\left( {\widehat {x}|\widehat {z}} \right)$ for
		the receiver with parameter set $\theta$.
		Mathematically, we aim to jointly optimize parameters $\phi$ and $\theta$ to maximize the log-likelihood of data
		$X$ as follows
		\begin{align}\label{LD1}
			\mathop {{\rm{max}}}\limits_{\phi ,\theta } \log {p_\theta }\left( {x} \right).
		\end{align}

		To efficiently handle optimization problem \eqref{LD1}, we optimize the lower bound of the objective function $\log {p_\theta }\left( {x} \right)$ \cite{Kingma_arXiv}.
		Specifically, $\log {p_\theta }\left( {x} \right)$ is lower bounded by
		\begin{subequations}
			\begin{align}
					\log {p_\theta }\left( {x} \right) &= \int\limits_{z} {{q_\phi }\left( {{z}|{x}} \right)\log {p_\theta }\left( {x} \right)} d{z}\label{LB_a}\\
				& = \int\limits_{z} {{q_\phi }\left( {{z}|{x}} \right)\log \frac{{{p_\theta }\left( {{z},{x}} \right)}}{{{p_\theta }\left( {{z}|{x}} \right)}}} d{z}\\
				& = \int\limits_{z} {{q_\phi }\left( {{z}|{x}} \right)\log \frac{{{p_\theta }\left( {{z},{x}} \right)}}{{{q_\phi }\left( {{z}|{x}} \right)}}} d{z} + \int\limits_{z} {{q_\phi }\left( {{z}|{x}} \right)\log \frac{{{q_\phi }\left( {{z}|{x}} \right)}}{{{p_\theta }\left( {{z}|{x}} \right)}}} d{z}\\
				& = \int\limits_{z} {{q_\phi }\left( {{z}|{x}} \right)\log \frac{{{p_\theta }\left( {{z},{x}} \right)}}{{{q_\phi }\left( {{z}|{x}} \right)}}} d{z} + {d_{{\rm{KL}}}}\left( {{q_\phi }\left( {{z}|{x}} \right)||{p_\theta }\left( {{z}|{x}} \right)} \right)\\
				&\ge \int\limits_{z} {{q_\phi }\left( {{z}|{x}} \right)\log \frac{{{p_\theta }\left( {{z},{x}} \right)}}{{q_\phi\left( {{z}|{x}} \right)}}} d{z},\label{LB_b}\\
				&= \int\limits_{z} {{q_\phi }\left( {{z}|{x}} \right)\log \frac{{{p_\theta }\left( {{x}|{z}} \right){p_\theta }\left( {z} \right)}}{{{q_\phi }\left( {{z}|{x}} \right)}}} d{z}\\
				&= \int\limits_{z} {{q_\phi }\left( {{z}|{x}} \right)\log {p_\theta }\left( {{x}|{z}} \right)}d{z} + \int\limits_{z} {{q_\phi }\left( {{z}|{x}} \right)\log \frac{{{p_\theta }\left( {z} \right)}}{{{q_\phi }\left( {{z}|{x}} \right)}}} d{z}\\
				&= {{\mathbb{E}}_{{q_\phi }\left( {{z}|{x}} \right)}}\left[ {\log {p_\theta }\left( {{x}|{z}} \right)} \right] - {d_{{\rm{KL}}}}\left( {{q_\phi }\left( {{z}|{x}} \right)||{p_\theta }\left( {z} \right)} \right)\label{LB_c}
			\end{align}
		\end{subequations}
	where equation \eqref{LB_a} holds for the arbitrary distribution ${{q_\phi }\left( {{z}|{x}} \right)}$, and inequality \eqref{LB_b} holds due to ${d_{{\rm{KL}}}}\left( {{q_\phi }\left( {{z}|{x}} \right)||{p_\theta }\left( {{z}|{x}} \right)} \right) \ge 0$.
			
		Unfortunately, maximizing the lower bound in \eqref{LB_c} directly cannot achieve interpretable and robust semantic communication systems design. To address this challenge, we multiply ${d_{{\rm{KL}}}}\left( {{q_\phi }\left( {{z}|{x}} \right)||{p_\theta }\left( {z} \right)} \right)$ by a weighting parameter $\beta $ to obtain a disentangling and explainable semantic representation ${z}$ \cite{Higgins_ICLR_2017}, for $\beta > 1$.
		Furthermore, to combat semantic noise and achieve robust semantic communication systems design, we replace ${{p_\theta }\left( {{z}|{x}} \right)}$ with ${p_\theta }\left( {x|\widehat z} \right)$, where $\widehat z = {\rm{g}}z + {n_{\rm{s}}}$, ${\rm{g}}$ denotes fading channel gain, and
${{{{n}}_{\rm{s}}}}$ denotes the semantic noise.
		Specifically, the log-likelihood maximization problem \eqref{LD1} is reformulated as follows
		\begin{align}\label{LD_vae}
			\mathop {{\rm{max}}}\limits_{\phi ,\theta } {{\mathbb{E}}_{{q_\phi }\left( {{x}|{z}} \right)}}\left[ \log {p_\theta }\left( {x|\widehat z} \right) \right] - \beta {d_{{\rm{KL}}}}\left( {{q_\phi }\left( {{z}|{x}} \right)||{p_\theta }\left( {z} \right)} \right),
		\end{align}
		where the prior distribution
		${p_\theta }\left( z \right) $ is assumed to follow a standard Gaussian distribution, i.e., ${p_\theta }\left( z \right) = \mathcal{N}\left( {{\bf{0}},{\bf{I}}} \right)$.Note that, in \eqref{LD_vae},
the first term ${{\mathbb{E}}_{{q_\phi }\left( {{x}|{z}} \right)}}\left[ {\log {p_\theta }\left( {{x}|{z} + {{{n}}_{\rm{s}}}} \right)} \right]$ is the expected likelihood with the cross entropy form, which can be regarded as reconstruction loss, while the second term
 regularizes ${q_\phi }\left( {{z}|{x}} \right)$ to be close to prior ${p_\theta }\left( {z} \right)$, which can be regarded as regularization loss.
To further enhance the robustness of the variational inference, we exploit $\eta $-cross entropy ${{c}_\eta }\left( {p\left( x \right)||{p_\theta }\left( {x|\hat z} \right)} \right)$ \cite{Eguchi_Entropy_2010,Futami_arxiv_2017} as the reconstruction loss, instead of
the cross entropy ${{\mathbb{E}}_{{q_\phi }\left( {{x}|{z}} \right)}}\left[ {\log {p_\theta }\left( {{x}|{z} + {{{n}}_{\rm{s}}}} \right)} \right]$,
where \begin{align}{{c}_\eta }\left( {p\left( x \right)||{p_\theta }\left( {x|\hat z} \right)} \right) = - \frac{{\eta + 1}}{\eta }\int {p{{\left( x \right)}^\eta }dx} + \int {{p_\theta }{{\left( {x|\hat z} \right)}^{1 + \eta }}dx}. \end{align}
Specifically, the objective function of the proposed robust semantic communication system is given as
 \begin{align}\label{Robust_VAE}\mathop {{\rm{max}}}\limits_{\phi ,\theta } {c_\eta }\left( {p\left( x \right)||{p_\theta }\left( {x|\hat z} \right)} \right) - \beta {d_{{\rm{KL}}}}\left( {{q_\phi }\left( {z|x} \right)||{p_\theta }\left( z \right)} \right).
 \end{align}

		Thus, the robust $\beta$-VAE training objective \eqref{Robust_VAE} encourages the latent distribution
		${{q_\phi }\left( {z|x} \right)}$ to efficiently represent semantic information about the data $x$
		by jointly maximizing the $\eta $-cross entropy ${{c}_\eta }\left( {p\left( x \right)||{p_\theta }\left( {x|\hat z} \right)} \right)$ and
		minimizing the $\beta$-weighted
		KL term ${d_{{\rm{KL}}}}\left( {{q_\phi }\left( {{z}|{x}} \right)||{p_\theta }\left( {z} \right)} \right)$ via unsupervised learning.
		
		More specifically,
		we jointly optimize the semantic encoder parameter $\phi$ and semantic
		decoder parameter $\theta$
		to maximize the objective function \eqref{Robust_VAE}.
		The first term of \eqref{Robust_VAE} is
		the probability of reconstructing the input data $x$,
		which corresponds to reconstruction loss. The second term is
		minimizing the KL divergence, which is the distance between
		the approximated posterior ${{q_\phi }\left( {z|x} \right)}$ and the fixed
		Gaussian distribution ${p_\theta }\left( z \right) = \mathcal{N}\left( {{\bf{0}},{\bf{I}}} \right)$.
		By adopting the well chosen values of the parameter $\beta $ (usually $\beta >1 $),
		the posterior ${{q_\phi }\left( {z|x} \right)}$ is encouraged to match the
		Gaussian distribution ${p_\theta }\left( z \right) = \mathcal{N}\left( {{\bf{0}},{\bf{I}}} \right)$, which
		disentangles
		the hidden representation into multiple independent and semantically meaningful
		features $\left\{ {{{z}_{l}}} \right\}_{l \in {\mathcal{L}}}$.
		The parameter $\beta$ balances reconstruction accuracy
		and learned disentanglement quality. In general, a higher value of $\beta$ will produce a more disentangled representation, but may lead to lower reconstruction
		accuracy\cite{Higgins_ICLR_2017}.
		
					\begin{figure*}[htbp]
			\centering
			\includegraphics[width=0.9\textwidth]{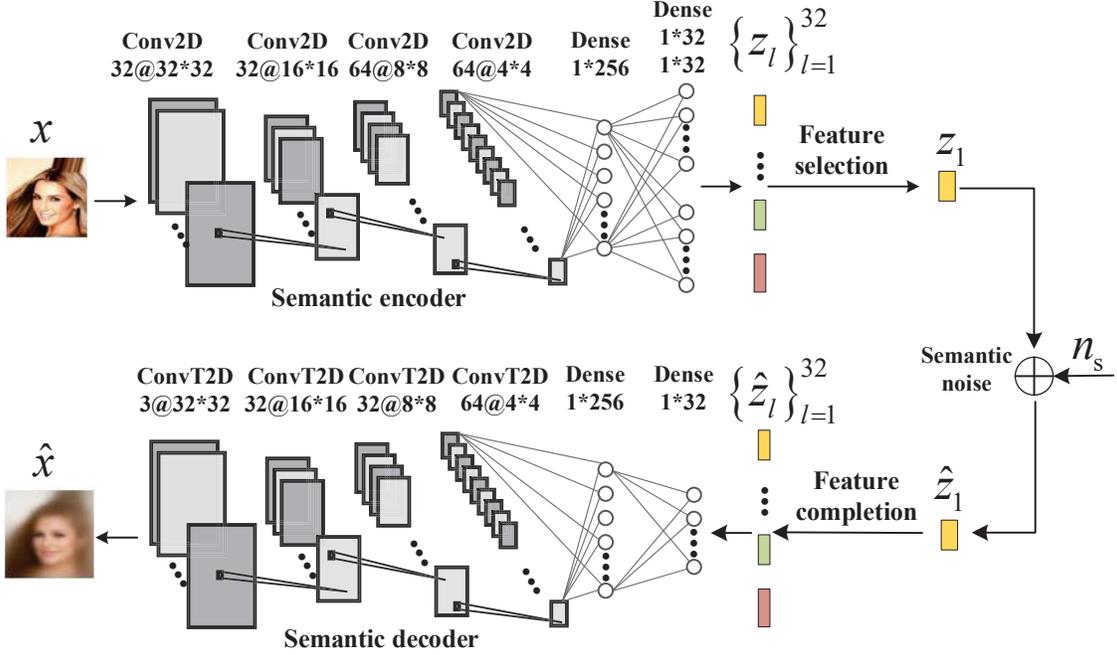}
			\caption{Proposed robust $\beta$-VAE based architecture for the explainable semantic communication system}
			\label{beta_SM1}
		\end{figure*}

		Note that, in the robust $\beta$-VAE network, we let $\left\{ {{\mu _l}} \right\}_{l = 1}^L$ and $\left\{ {{\sigma _l}} \right\}_{l = 1}^L$ denote the mean and the corresponding standard
		deviation of the approximate posterior ${q_\phi }\left( {{z}|{x}} \right)$, respectively.
		Moreover, a reparametrization trick \cite{Higgins_ICLR_2017} is applied to estimate
		gradients of the objective function \eqref{LD_vae} with respect to the parameter $\phi$,
		where random independent variables $\left\{ {{\varepsilon _l}} \right\}_{l = 1}^L$ are sampled from a standard Gaussian distribution, i.e., ${{\varepsilon _l}} \sim \mathcal{N}\left( {{{0}},{\bf{1}}} \right)$.
		Then, the output features of the semantic encoder $\left\{ {{z_l}} \right\}_{l = 1}^L$ are given as follows
		\begin{align}{z_l} = {\mu _l} + {\sigma _l}{\varepsilon _l},~l = 1,...,L.
		\end{align}
		Thus, the feature $z_l$ is equivalent to being sampled from distribution
		${\cal N}\left( {{\mu _l},\sigma _l^2} \right)$, where $l = 1,...,L$.

		
		\subsection{ Feature Selection and Completion}
		
		With the disentangled and explainable features, the proposed semantic communications system further performs feature selection and completion at the transmitter and receiver, respectively.
		Specifically, since the receiver may only be interested in some of the features,
		the transmitter only sends the intended features ${\left\{ {{{z}_{l}}} \right\}_{l \in {\mathcal{L} _{\rm{sel}}}}}$ according to their semantic meanings,
		rather than all of the extracted features $\left\{ {{{z}_{l}}} \right\}_{l \in {\mathcal{L}}}$,
		which can further reduce the amount of information transmission.

		For the receiver, the proposed semantic source and channel decoder include semantic feature completion and feature reconstruction. Specifically,
		for the unintended features subset are not transmitted ${\left\{ {{z_l}} \right\}_{l \in \mathcal{L}\backslash {\mathcal{L}_{{\rm{sel}}}}}}$, the receiver
		generated the corresponding features ${\left\{ {{{\widehat {z}}_{l}}} \right\}_{l \in \mathcal{L}\backslash {\mathcal{L}_{{\rm{sel}}}}}}$ based on the receiver knowledge base, where both the dimensions and value ranges of sets ${z_l}$ and ${{{\widehat {z}}_{l}}}$ are the same.
		
		Then, according to the completed semantic features $\widehat Z = {\left\{ {{{\hat z}_l}} \right\}_{l \in \mathcal{L}}}$, the feature reconstruction module recovers the original data $\widehat X$.

		\subsection{Proposed Architecture}

		The proposed lightweight semantic communication architecture includes a semantic encoder network and a semantic decoder network, as shown in Fig. \ref{beta_SM1}, where the notation Conv2D
		32$@$32*32 means that the network has 32
		2-D convolutional filters of size 32*32, and Dense
		1*256
		represents a dense layer with 256 neurons.
		The details of the semantic encoder and the decoder network architectures
		are given as:
		
		\subsubsection{Semantic encoder architecture}: Conv2D
		32$@$32*32 $\to$ Conv2D
		32$@$16*16
		$\to$ Conv2D
		64$@$8*8 $\to$ Conv2D
		64$@$4*4 $\to$ Dense
		1*256 $\to$
		2 parallel Dense
		1*32
		$\to$ $\left\{ {{z_l}} \right\}_{l = 1}^{32}$ $\to$ ${\left\{ {{{z}_{l}}} \right\}_{l \in {\mathcal{L} _{\rm{sel}}}}}$;
		
		\subsubsection{Semantic decoder architecture}:
		${\left\{ {{{\widehat z}_l}} \right\}_{l \in {{\cal L}_{{\rm{sel}}}}}}$ $\to$ $\left\{ {{{\widehat z}_l}} \right\}_{l = 1}^{32}$ $\to$
		Dense
		1*32 $\to$ Dense
		1*256
		$\to$ ConvT2D
		64$@$4*4 $\to$ ConvT2D
		32$@$8*8 $\to$ ConvT2D
		32$@$16*16 $\to$ ConvT2D
		3$@$32*32.
		
		Note that, based on the feature selection,
		the proposed semantic communication system only needs to send the features ${\left\{ {{{z}_{l}}} \right\}_{l \in {\mathcal{L} _{\rm{sel}}}}}$ that the receiver is interested in, instead of sending all features $\left\{ {{{ z}_l}} \right\}_{l = 1}^{32}$.
		
		\section{Prototype and Implementations}

\begin{figure}[htbp]
			\centering	
			\includegraphics[width=0.7\textwidth]{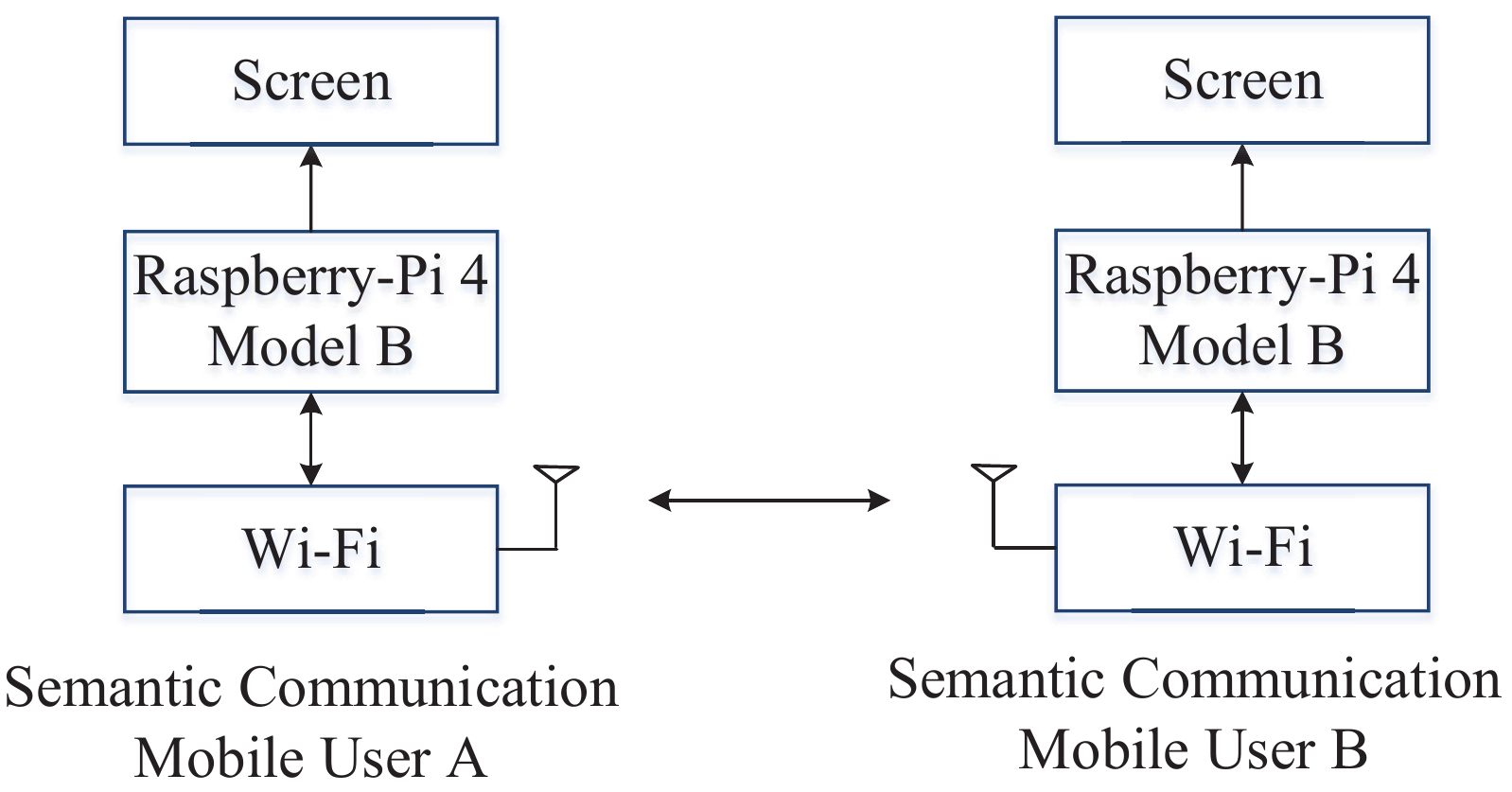}
			\vskip0.1cm\centering {\footnotesize (a)}
			
			\centering
			\includegraphics[width=0.7\textwidth]{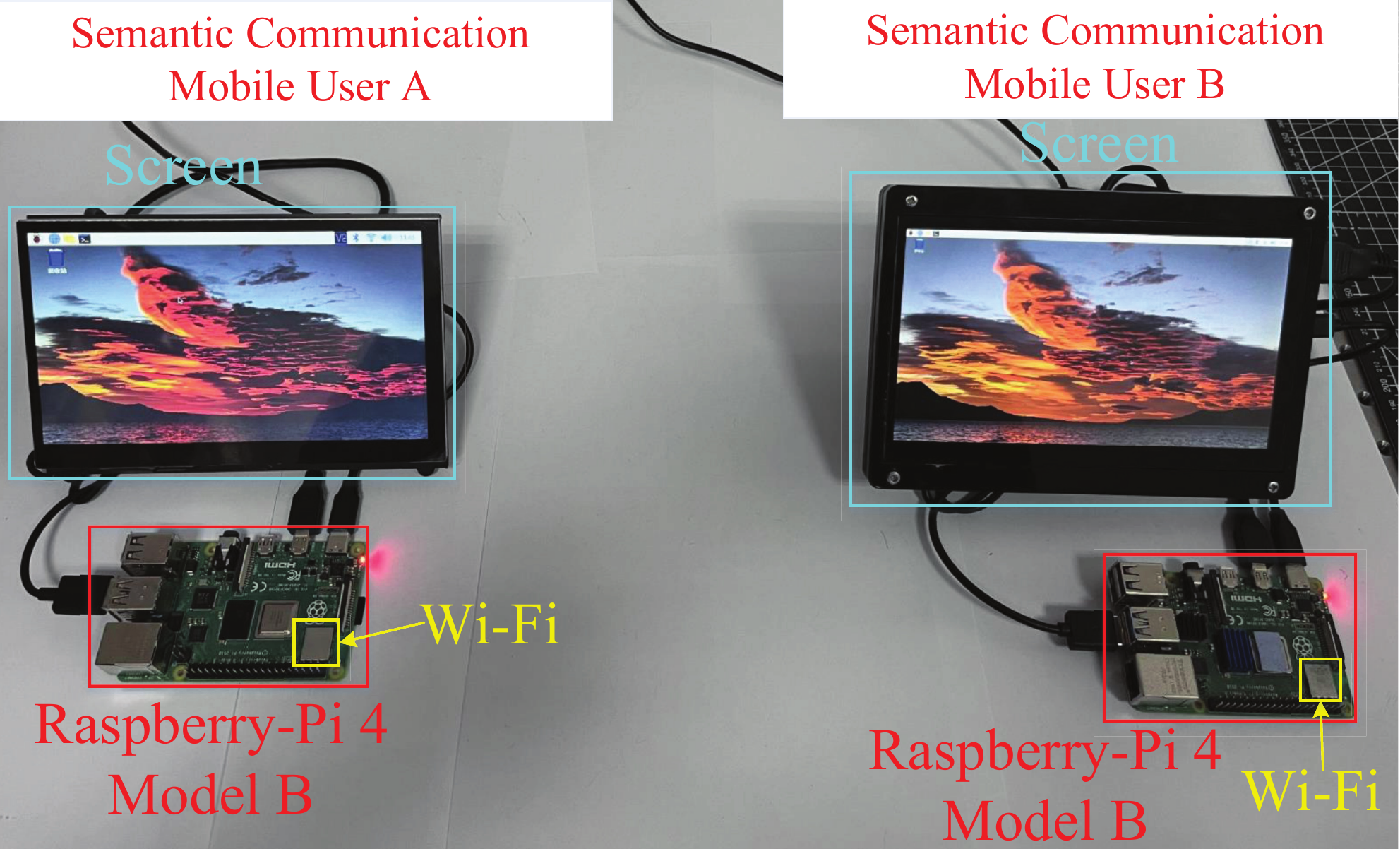}
			\vskip0.1cm\centering {\footnotesize (b)}
			
			\caption{
				(a) The architecture of the semantic
				communication system prototype;
				(b) The hardware platform of the semantic communication system prototype.}
			\label{ESM prototype}
		\end{figure}

		The proposed architecture and hardware platform design of
		the semantic communication system prototype are shown in Fig. \ref{ESM prototype} (a) and (b), which can be used to implement the proposed robust and explainable semantic
		communications system in Fig. \ref{beta_SM}.		
		The prototype includes two semantic communication
		mobile users A and B.
		The trained robust $\beta$-VAE network is implemented at the portable RaspberryPi 4 Model B
		processors to realize the semantic encoding/decoding and the feature selection/completion functions. The integrated Wi-Fi module fulfills the bit-level transmission. The decoded data can be shown through the display.

		The detailed parameters of the prototype are provided in Table \ref{canshu}.
		The Raspberry Pi is installed with an ARM Cortex-A72$@$quad-core 1.5GHz CPU and 4GB of DDR4 RAM, and is equipped with Pytorch-CPU and torchvision software. The communication between Raspberry Pi A and B is realized through WiFi, where the socket is used to send and receive data, and Visdom is used to realize visual communication.

		%

		\begin{table}[!hbp]
			\centering
			\caption{Hardware parameters of the semantic
				communication prototype.}
			\begin{tabular}{l|l}
				\hline
				GPU & 500MHz VideoCore VI \\
				\hline
				CPU& quad-core Cortex-A72
				\\
				\hline
				System on Chip & Broadcom BCM2711$@$ 1.5GHz\\
				\hline
				memory & 4GB DDR4\\
				\hline
				Wi-Fi & 2.4$/$ 5.0 GHz IEEE 802.11ac wireless\\
				\hline
				Screen & $800 \times 480$ display\\
				\hline
			\end{tabular}\label{canshu}
		\end{table}

		\section{ Experiments and Discussions}
		
		In this section, we evaluate the proposed
		explainable semantic communications system using a graphics processing unit (GPU) and
		Raspberry Pi prototype, respectively.
		The GPU experiments in this work have been performed on 32 GB RAM i5-12600H,
		and 8 GB Nvidia GeForce 3060Ti GTX graphics card with Pytorch powered with CUDA 11.3.
		The experiments are performed via two standard
		datasets, i.e., MNIST Dataset and CelebA Dataset.

		\subsection{Demonstration via GPU}

		First, we evaluate the robustness of the proposed semantic communication system.
 Specifically, the peak signal-to-noise
		ratio (PSNR) performance of
 the proposed robust $\beta$-VAE scheme with ${\rm{SNR}}_{{\rm{train}}}= 4{\rm{dB}}$ and ${\rm{SNR}}_{{\rm{train}}}= 8{\rm{dB}}$ are
 demonstrated over the two channel models: the ANGC and a slow Rayleigh fading channel,
		where ${\rm{SNR}}_{{\rm{train}}}= 4{\rm{dB}}$ and ${\rm{SNR}}_{{\rm{train}}}= 8{\rm{dB}}$ mean that the
		trained SNRs of the schemes are
		$4{\rm{dB}}$ and $8{\rm{dB}}$, respectively.
 Moreover,  the PSNR  performance of the deep joint source-channel coding (Deep-JSCC) scheme \cite{Bourtsoulatze_TCCN_2019},    $\beta$-VAE scheme, and the JPEG compression scheme
  are  presented for comparisons.

		\begin{figure*}[htbp]
			\centering
			\begin{minipage}[t]{0.46\textwidth}
				\centering
				\includegraphics[width=\textwidth]{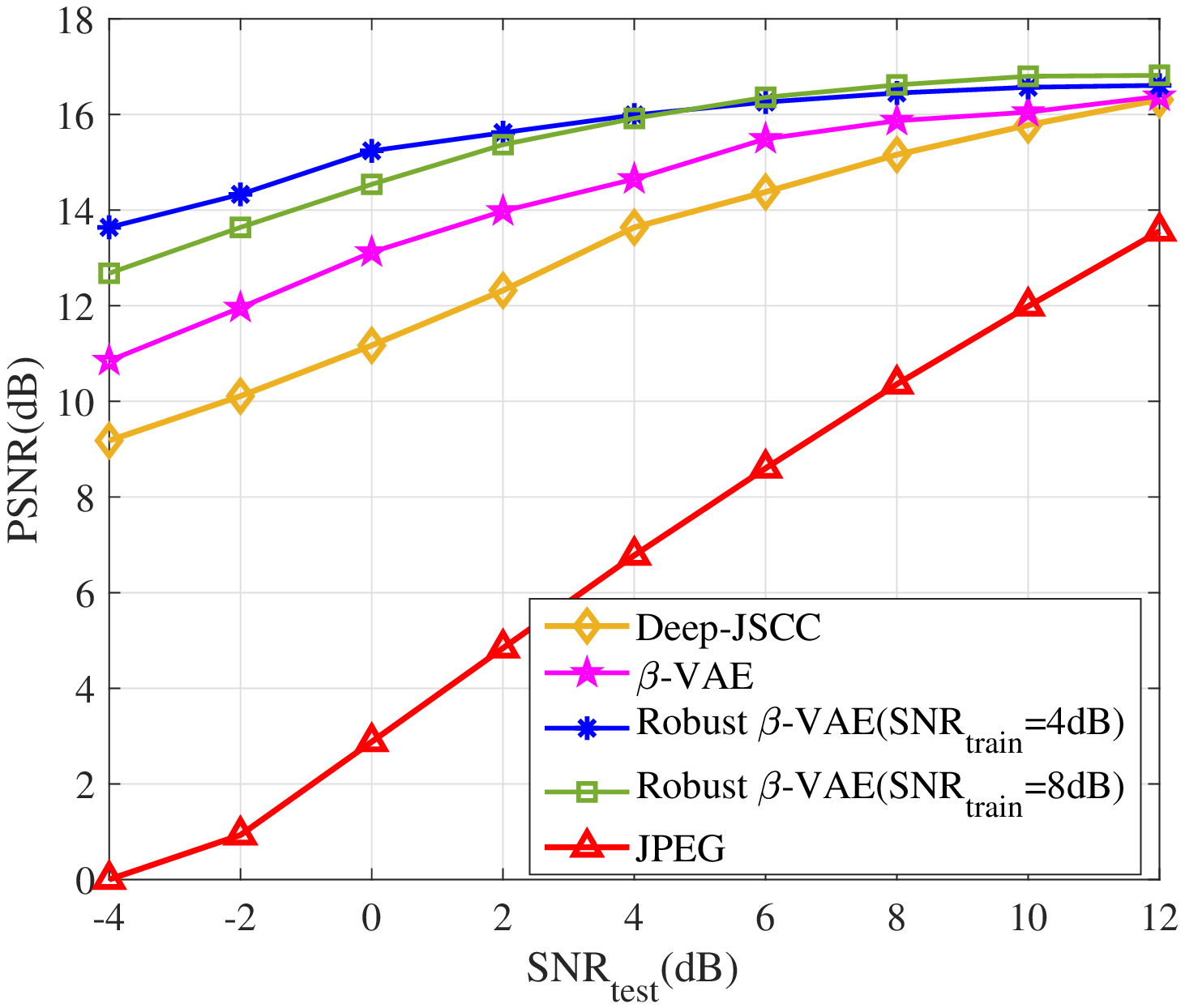}
				\vskip-0.2cm\centering {\footnotesize (a)}
			\end{minipage}
			\begin{minipage}[t]{0.48\textwidth}
				\centering
				\includegraphics[width=\textwidth]{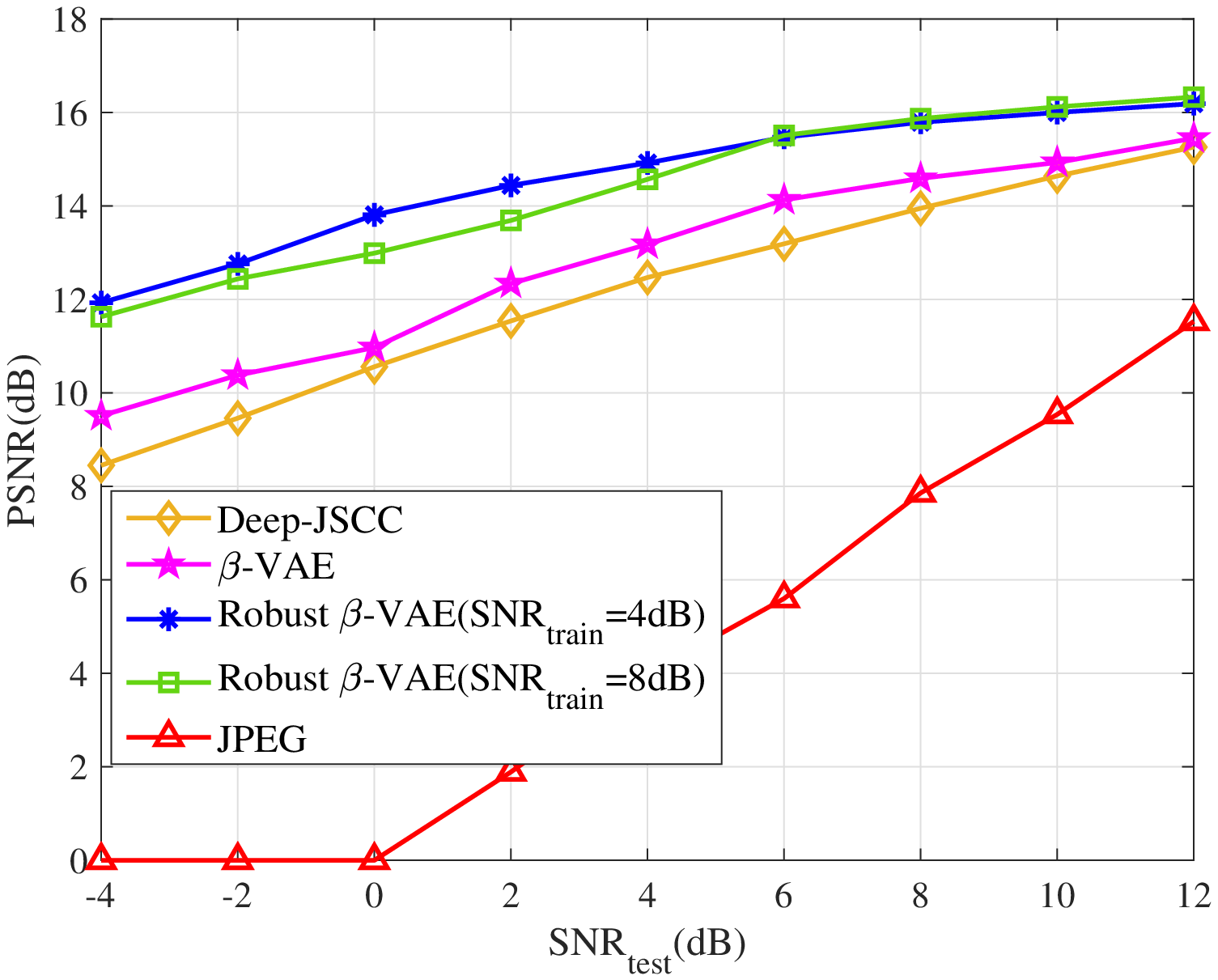}
				\vskip-0.2cm\centering {\footnotesize (b)}
			\end{minipage}
			\caption{ (a) PSNR of Deep-JSCC, $\beta$-VAE, JPEG compression,  and robust $\beta$-VAE with ${\rm{SNR}}_{{\rm{train}}}= 4{\rm{dB}}$ and ${\rm{SNR}}_{{\rm{train}}}= 8{\rm{dB}}$ over ANGC with semantic noise $a=-0.1$, $b=0.1$ and $\sigma _{\rm{P}}^2 = 1$;
				(b) PSNR of Deep-JSCC,  $\beta$-VAE, JPEG compression, and robust $\beta$-VAE with ${\rm{SNR}}_{{\rm{train}}}= 4{\rm{dB}}$ and ${\rm{SNR}}_{{\rm{train}}}= 8{\rm{dB}}$ over Rayleigh fading channels with $\sigma _{\rm{h}}^2 = 1$, $a=-0.1$, $b=0.1$ and $\sigma _{\rm{P}}^2 = 1$.}
			\label{ANGC_Rayleigh}
		\end{figure*}
		Fig. \ref{ANGC_Rayleigh} (a) shows PSNR versus different test SNRs of the four schemes over ANGC, where semantic noise parameters $a=-0.1$, $b=0.1$ and $\sigma _{\rm{P}}^2 = 1$.
	We observe that the PSNR of JPEG compression is the lowest among the five schemes, and the PSNR of the robust $\beta$-VAE schemes are higher than those  of both Deep-JSCC and $\beta$-VAE.
		In the low SNR regions, the PSNR of the robust $\beta$-VAE with ${\rm{SNR}}_{{\rm{train}}}= 4{\rm{dB}}$ is the highest, and the PSNR of the robust $\beta$-VAE with ${\rm{SNR}}_{{\rm{train}}}= 8{\rm{dB}}$ is the higher than that of $\beta$-VAE,
		which verifies the robustness of our proposed design. Since the training noise of ${\rm{SNR}}_{{\rm{train}}}= 4{\rm{dB}}$ is higher than that of ${\rm{SNR}}_{{\rm{train}}}= 8{\rm{dB}}$, the performance of ${\rm{SNR}}_{{\rm{train}}}= 4{\rm{dB}}$ is more robust, and thus
		the PSNR of ${\rm{SNR}}_{{\rm{train}}}= 4{\rm{dB}}$ is higher.
		In the high SNR regions, the PSNR of $\beta$-VAE, and robust $\beta$-VAE models tend to be the same.
		The reason is that the effect of noise at high SNR can be ignored.


		Fig. \ref{ANGC_Rayleigh} (b) illustrates PSNR versus different test SNRs of the five schemes over the Rayleigh fading channel.
		Similar to Fig. \ref{ANGC_Rayleigh} (a), the PSNR of JPEG compression is the lowest among the four schemes, and the PSNR of the robust $\beta$-VAE with ${\rm{SNR}}_{{\rm{train}}}= 4{\rm{dB}}$ and ${\rm{SNR}}_{{\rm{train}}}= 8{\rm{dB}}$ are higher than     those  of both Deep-JSCC and $\beta$-VAE.
		Note that for ${\rm{SN}}{{\rm{R}}_{{\rm{test}}}} = 8{\rm{dB}}$, the PSNR of the robust $\beta$-VAE with ${\rm{SNR}}_{{\rm{train}}}= 8{\rm{dB}}$ is the higher than that of ${\rm{SNR}}_{{\rm{train}}}= 8{\rm{dB}}$. This because the training SNR of ${\rm{SNR}}_{{\rm{train}}}= 8{\rm{dB}}$ is also $8$dB.
		Comparing Fig. \ref{ANGC_Rayleigh} (a) with
		ANGC, the PSNRs of the schemes in Fig. \ref{ANGC_Rayleigh} (b) are lower due to Rayleigh random fading.

		\begin{table*}[htbp]
			\caption{Transmission performance comparison over ANGC}
			\centering
			\begin{tabular}{ | c | c | c | c | c|c|}
				\hline
				Test SNR & Data $X$ & JPEG &\tabincell{c}{$\beta$-VAE} &\tabincell{c}{Robust $\beta$-VAE \\${\rm{SNR}}_{{\rm{train}}}= 4{\rm{dB}}$}&\tabincell{c}{Robust $\beta$-VAE \\${\rm{SNR}}_{{\rm{train}}}= 8{\rm{dB}}$ } \\ \hline
				SNR\_test=4dB&\begin{minipage}[b]{0.1\columnwidth}
					\centering
					\raisebox{-.5\height}{\includegraphics[width=\linewidth]{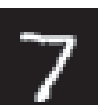}}
				\end{minipage}&\begin{minipage}[b]{0.1\columnwidth}
					\centering
					 \raisebox{-.5\height}{\includegraphics[width=\linewidth]{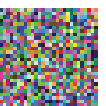}}
				\end{minipage}
				&\begin{minipage}[b]{0.1\columnwidth}
					\centering
					 \raisebox{-.5\height}{\includegraphics[width=\linewidth]{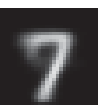}}
				\end{minipage}
				&\begin{minipage}[b]{0.1\columnwidth}
					
					\centering
					 \raisebox{-.5\height}{\includegraphics[width=\linewidth]{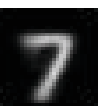}}
				\end{minipage}
				&\begin{minipage}[b]{0.1\columnwidth}
					\centering
					 \raisebox{-.5\height}{\includegraphics[width=\linewidth]{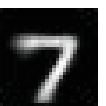}}
				\end{minipage}

				\\ \hline
				
				SNR\_test=8dB&\begin{minipage}[b]{0.1\columnwidth}
					\centering
					\raisebox{-.5\height}{\includegraphics[width=\linewidth]{x.eps}}
				\end{minipage}&\begin{minipage}[b]{0.1\columnwidth}
					\centering
					 \raisebox{-.5\height}{\includegraphics[width=\linewidth]{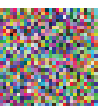}}
				\end{minipage}
				&\begin{minipage}[b]{0.1\columnwidth}
					\centering
					 \raisebox{-.5\height}{\includegraphics[width=\linewidth]{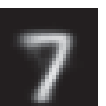}}
				\end{minipage}
				&\begin{minipage}[b]{0.1\columnwidth}
					\centering
					 \raisebox{-.5\height}{\includegraphics[width=\linewidth]{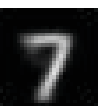}}
				\end{minipage}
				&\begin{minipage}[b]{0.1\columnwidth}
					\centering
					 \raisebox{-.5\height}{\includegraphics[width=\linewidth]{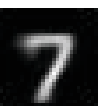}}
				\end{minipage}
				
				\\ \hline
				SNR\_test=4dB&\begin{minipage}[b]{0.1\columnwidth}
					\centering
					\raisebox{-.5\height}{\includegraphics[width=\linewidth]{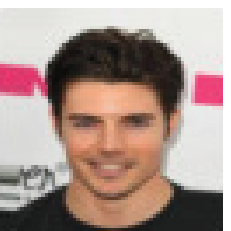}}
				\end{minipage}&\begin{minipage}[b]{0.1\columnwidth}
					\centering
					 \raisebox{-.5\height}{\includegraphics[width=\linewidth]{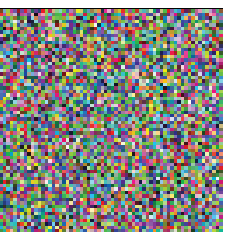}}
				\end{minipage}
				&\begin{minipage}[b]{0.1\columnwidth}
					\centering
					 \raisebox{-.5\height}{\includegraphics[width=\linewidth]{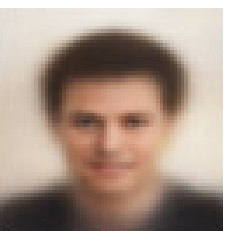}}
				\end{minipage}
				&\begin{minipage}[b]{0.1\columnwidth}
					\centering
					 \raisebox{-.5\height}{\includegraphics[width=\linewidth]{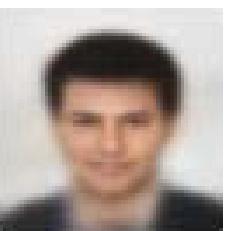}}
				\end{minipage}
				&\begin{minipage}[b]{0.1\columnwidth}
					\centering
					 \raisebox{-.5\height}{\includegraphics[width=\linewidth]{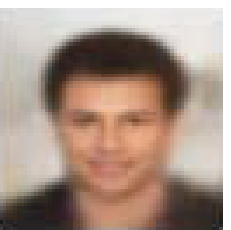}}
				\end{minipage}
				
				\\ \hline
				
				SNR\_test=8dB&\begin{minipage}[b]{0.1\columnwidth}
					\centering
					\raisebox{-.5\height}{\includegraphics[width=\linewidth]{x_1.eps}}
				\end{minipage}&\begin{minipage}[b]{0.1\columnwidth}
					\centering
					 \raisebox{-.5\height}{\includegraphics[width=\linewidth]{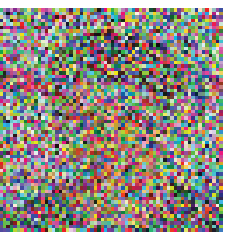}}
				\end{minipage}
				&\begin{minipage}[b]{0.1\columnwidth}
					\centering
					 \raisebox{-.5\height}{\includegraphics[width=\linewidth]{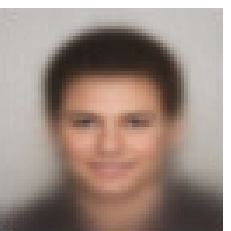}}
				\end{minipage}
				&\begin{minipage}[b]{0.1\columnwidth}
					\centering
					 \raisebox{-.5\height}{\includegraphics[width=\linewidth]{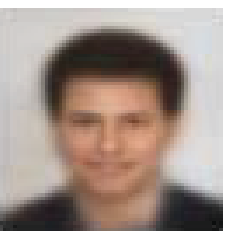}}
				\end{minipage}
				&\begin{minipage}[b]{0.1\columnwidth}
					\centering
					 \raisebox{-.5\height}{\includegraphics[width=\linewidth]{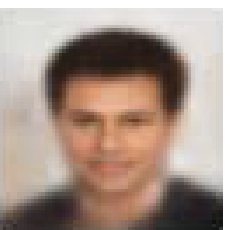}}
				\end{minipage}
				
				\\ \hline
				
			\end{tabular}\label{Transmission_ANGC}
		\end{table*}

		\begin{table*}[ht]
			\caption{Transmission performance comparison over Rayleigh fading channel}
			\centering
			\begin{tabular}{ | c | c | c | c | c|c|}
				\hline
				Test SNR & Data $X$ & JPEG &\tabincell{c}{$\beta$-VAE} &\tabincell{c}{Robust $\beta$-VAE \\${\rm{SNR}}_{{\rm{train}}}= 4{\rm{dB}}$}&\tabincell{c}{Robust $\beta$-VAE \\${\rm{SNR}}_{{\rm{train}}}= 8{\rm{dB}}$ } \\ \hline
				SNR\_test=4dB&\begin{minipage}[b]{0.1\columnwidth}
					\centering
					\raisebox{-.5\height}{\includegraphics[width=\linewidth]{x.eps}}
				\end{minipage}&\begin{minipage}[b]{0.1\columnwidth}
					\centering
					 \raisebox{-.5\height}{\includegraphics[width=\linewidth]{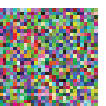}}
				\end{minipage}
				&\begin{minipage}[b]{0.1\columnwidth}
					\centering
					 \raisebox{-.5\height}{\includegraphics[width=\linewidth]{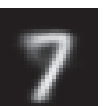}}
				\end{minipage}
				&\begin{minipage}[b]{0.1\columnwidth}
					\centering
					 \raisebox{-.5\height}{\includegraphics[width=\linewidth]{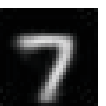}}
				\end{minipage}
				&\begin{minipage}[b]{0.1\columnwidth}
					\centering
					 \raisebox{-.5\height}{\includegraphics[width=\linewidth]{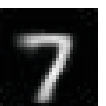}}
				\end{minipage}
				
				\\ \hline
				
				SNR\_test=8dB&\begin{minipage}[b]{0.1\columnwidth}
					\centering
					\raisebox{-.5\height}{\includegraphics[width=\linewidth]{x.eps}}
				\end{minipage}&\begin{minipage}[b]{0.1\columnwidth}
					\centering
					 \raisebox{-.5\height}{\includegraphics[width=\linewidth]{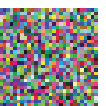}}
				\end{minipage}
				&\begin{minipage}[b]{0.1\columnwidth}
					\centering
					 \raisebox{-.5\height}{\includegraphics[width=\linewidth]{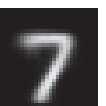}}
				\end{minipage}
				&\begin{minipage}[b]{0.1\columnwidth}
					\centering
					 \raisebox{-.5\height}{\includegraphics[width=\linewidth]{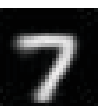}}
				\end{minipage}
				&\begin{minipage}[b]{0.1\columnwidth}
					\centering
					 \raisebox{-.5\height}{\includegraphics[width=\linewidth]{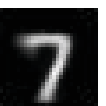}}
				\end{minipage}

				\\ \hline
				SNR\_test=4dB&\begin{minipage}[b]{0.1\columnwidth}
					\centering
					\raisebox{-.5\height}{\includegraphics[width=\linewidth]{x_1.eps}}
				\end{minipage}&\begin{minipage}[b]{0.1\columnwidth}
					\centering
					 \raisebox{-.5\height}{\includegraphics[width=\linewidth]{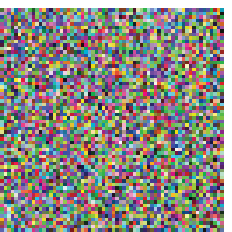}}
				\end{minipage}
				&\begin{minipage}[b]{0.1\columnwidth}
					\centering
					 \raisebox{-.5\height}{\includegraphics[width=\linewidth]{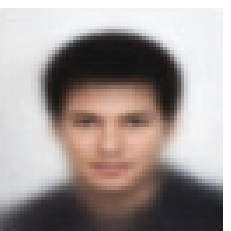}}
				\end{minipage}
				&\begin{minipage}[b]{0.1\columnwidth}
					\centering
					 \raisebox{-.5\height}{\includegraphics[width=\linewidth]{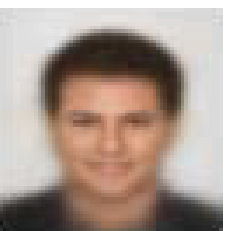}}
				\end{minipage}
				&\begin{minipage}[b]{0.1\columnwidth}
					\centering
					 \raisebox{-.5\height}{\includegraphics[width=\linewidth]{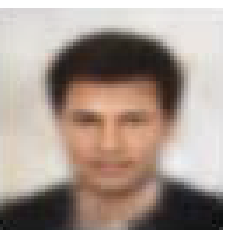}}
				\end{minipage}

				\\ \hline
				
				SNR\_test=8dB&\begin{minipage}[b]{0.1\columnwidth}
					\centering
					\raisebox{-.5\height}{\includegraphics[width=\linewidth]{x_1.eps}}
				\end{minipage}&\begin{minipage}[b]{0.1\columnwidth}
					\centering
					 \raisebox{-.5\height}{\includegraphics[width=\linewidth]{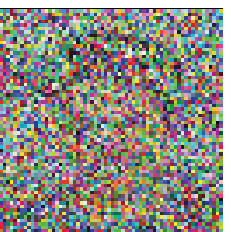}}
				\end{minipage}
				&\begin{minipage}[b]{0.1\columnwidth}
					\centering
					 \raisebox{-.5\height}{\includegraphics[width=\linewidth]{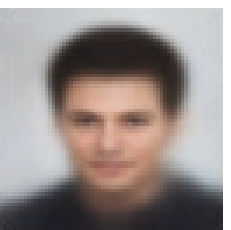}}
				\end{minipage}
				&\begin{minipage}[b]{0.1\columnwidth}
					\centering
					 \raisebox{-.5\height}{\includegraphics[width=\linewidth]{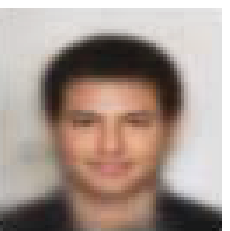}}
				\end{minipage}
				&\begin{minipage}[b]{0.1\columnwidth}
					\centering
					 \raisebox{-.5\height}{\includegraphics[width=\linewidth]{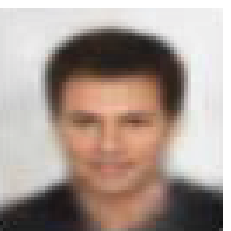}}
				\end{minipage}

				\\ \hline
				
			\end{tabular}\label{Transmission_Rayleigh}
		\end{table*}

		Table \ref{Transmission_ANGC} illustrates the transmission performance of
		JPEG compression, $\beta$-VAE, and robust $\beta$-VAE with ${\rm{SNR}}_{{\rm{train}}}= 4{\rm{dB}}$ and ${\rm{SNR}}_{{\rm{train}}}= 8{\rm{dB}}$ over ANGC with semantic noise parameters $a=-0.1$, $b=0.1$ and $\sigma _{\rm{P}}^2 = 1$. The second column of Table \ref{Transmission_ANGC} shows the transmission performance of the
		JPEG compression scheme, where the transmitted semantics cannot be recognized from the received image.
		The third column shows the results of the $\beta$-VAE
		scheme, where the transmission semantics can be recognized from the received image.
		The fourth and fifth columns show received images of the robust $\beta$-VAE
		scheme with ${\rm{SNR}}_{{\rm{train}}}= 4{\rm{dB}}$ and ${\rm{SNR}}_{{\rm{train}}}= 8{\rm{dB}}$,
		and the quality is better than that of the $\beta$-VAE
		scheme.

		Table \ref{Transmission_Rayleigh} illustrates transmission performance of
		the four schemes over the Rayleigh fading channel with semantic noise parameters $a=-0.1$, $b=0.1$ and $\sigma _{\rm{P}}^2 = 1$. The second column of table \ref{Transmission_ANGC} shows the transmission performance of the
		JPEG compression scheme, where the transmission semantics cannot be recognized from the received image.
		The third column shows the transmission performance of the $\beta$-VAE
		scheme, where the transmission semantics can be recognized from the received image.
		The fourth and fifth columns show the transmission performance of the robust $\beta$-VAE scheme with ${\rm{SNR}}_{{\rm{train}}}= 4{\rm{dB}}$ and ${\rm{SNR}}_{{\rm{train}}}= 8{\rm{dB}}$ , and the received image is better than that of the $\beta$-VAE
		scheme.
		
		\subsection{Demonstration via Prototype }

		\begin{table*}[ht]
			\caption{Proposed semantic communication with feature selection}
			\centering
			\begin{tabular}{ | c | c | c | c | c|}
				\hline
				\tabincell{c}{Intended\\ Feature} & Skin color &\tabincell{c}{Face\\ orientation} &Gender&Hairstyle \\ \hline
				\tabincell{c}{Source\\ data $X$}
				
				& \begin{minipage}[b]{0.1\columnwidth}
					\centering
					\raisebox{-.5\height}{\includegraphics[width=\linewidth]{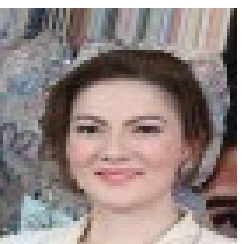}}
				\end{minipage}
				& \begin{minipage}[b]{0.1\columnwidth}
					\centering
					\raisebox{-.5\height}{\includegraphics[width=\linewidth]{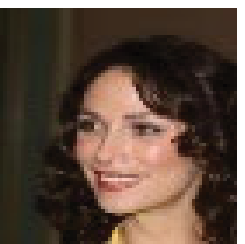}}
				\end{minipage}
				&\begin{minipage}[b]{0.1\columnwidth}
					\centering
					\raisebox{-.5\height}{\includegraphics[width=\linewidth]{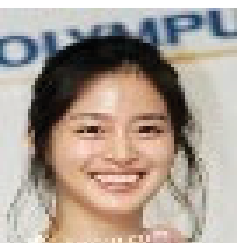}}
				\end{minipage}
				&\begin{minipage}[b]{0.1\columnwidth}
					\centering
					\raisebox{-.5\height}{\includegraphics[width=\linewidth]{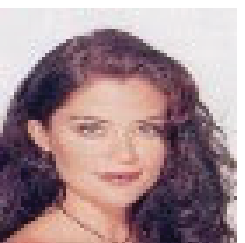}}
				\end{minipage}
				\\ \hline
				
				\tabincell{c}{Receiver\\ Knowledge\\ Base}

				& \begin{minipage}[b]{0.1\columnwidth}
					\centering
					\raisebox{-.5\height}{\includegraphics[width=\linewidth]{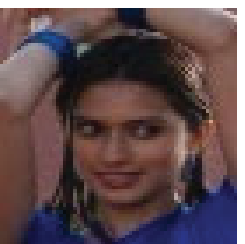}}
				\end{minipage}
				& \begin{minipage}[b]{0.1\columnwidth}
					\centering
					\raisebox{-.5\height}{\includegraphics[width=\linewidth]{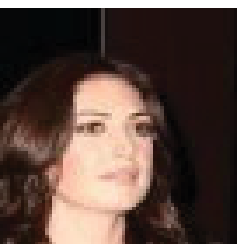}}
				\end{minipage}
				&\begin{minipage}[b]{0.1\columnwidth}
					\centering
					\raisebox{-.5\height}{\includegraphics[width=\linewidth]{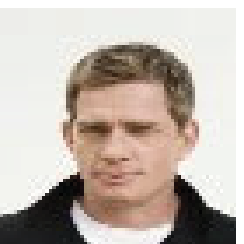}}
				\end{minipage}
				&\begin{minipage}[b]{0.1\columnwidth}
					\centering
					\raisebox{-.5\height}{\includegraphics[width=\linewidth]{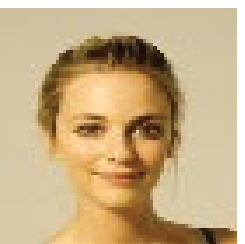}}
				\end{minipage}
				\\ \hline
				\tabincell{c}{Decoded\\ data $\hat X$ \\ (GPU)}
				
				& \begin{minipage}[b]{0.1\columnwidth}
					\centering
					\raisebox{-.5\height}{\includegraphics[width=\linewidth]{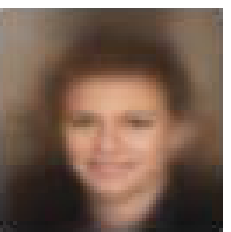}}
				\end{minipage}
				& \begin{minipage}[b]{0.1\columnwidth}
					\centering
					\raisebox{-.5\height}{\includegraphics[width=\linewidth]{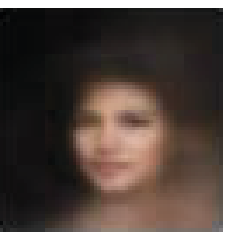}}
				\end{minipage}
				&\begin{minipage}[b]{0.1\columnwidth}
					\centering
					\raisebox{-.5\height}{\includegraphics[width=\linewidth]{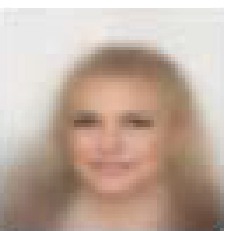}}
				\end{minipage}
				&\begin{minipage}[b]{0.1\columnwidth}
					\centering
					\raisebox{-.5\height}{\includegraphics[width=\linewidth]{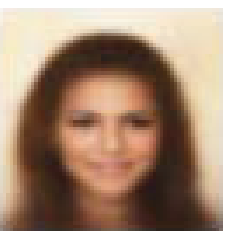}}
				\end{minipage}
				\\ \hline
				\tabincell{c}{Decoded\\ data $\hat X$ \\ (Raspberry)}
				
				& \begin{minipage}[b]{0.1\columnwidth}
					\centering
					\raisebox{-.5\height}{\includegraphics[width=\linewidth]{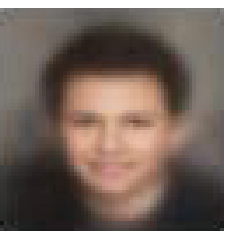}}
				\end{minipage}
				& \begin{minipage}[b]{0.1\columnwidth}
					\centering
					\raisebox{-.5\height}{\includegraphics[width=\linewidth]{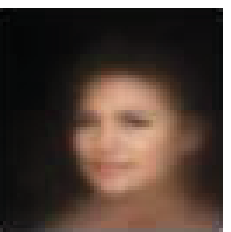}}
				\end{minipage}
				&\begin{minipage}[b]{0.1\columnwidth}
					\centering
					\raisebox{-.5\height}{\includegraphics[width=\linewidth]{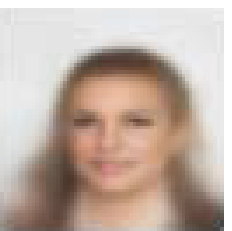}}
				\end{minipage}
				&\begin{minipage}[b]{0.1\columnwidth}
					\centering
					\raisebox{-.5\height}{\includegraphics[width=\linewidth]{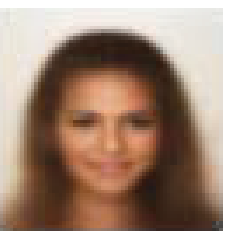}}
				\end{minipage}
				\\ \hline
			\end{tabular}\label{feature_sel}
		\end{table*}

		
		In this subsection, we demonstrate that the proposed explainable semantic communication system with
		feature selection can improve the transmission
		efficiency via our prototype.
		
		Table \ref{feature_sel} shows the performance of the proposed explainable semantic communication system
		with feature selection.
		From Column 2 to Column 5, we present four examples to show how the explainable encoder and feature selection work in the transmission.
		In the second column, the intended feature to send is skin color. The proposed semantic communication system performs feature extraction on the input white-skinned women picture, and then only selects the white skin color feature for transmission. Although the woman in the receiving knowledge base has darker skin, the reconstructed image is changed to white-skin.
		In the third column, the intended feature is face orientation.
		The proposed semantic communication system can successfully reconstruct a picture with the same face orientation at the receiver.
		Similarly, the intended features of the third and fourth columns are gender and hairstyle, respectively,
		and the proposed semantic communication system can also recover the correct feature at the receiver.
		
		Table \ref{prototype_MNIST} compares a compression ratio, transmission time, PSNR and reconstructed image of the original image transmission scheme, JPEG compression scheme, $\beta$-VAE scheme, and robust $\beta$-VAE scheme over our proposed semantic communication prototype on MNIST dataset with high SNR.
		From Table \ref{prototype_MNIST}, we observe that the compression ratio of the $\beta$-VAE scheme and robust $\beta$-VAE scheme is $78.4$ which is significantly higher than those of the JPEG compression scheme (1.81) and original image transmission scheme. Thus, the transmission time of the $\beta$-VAE scheme and robust $\beta$-VAE scheme is about $0.3$ms, which is significantly lower than those of the JPEG compression scheme (10.88ms) and original image transmission scheme (18.86ms). Therefore, the proposed semantic communication system can significantly reduce the transmission load and time.
		Moreover, the PSNR of the robust $\beta$-VAE scheme is close to that of the JPEG compression scheme, and is higher than that of the $\beta$-VAE scheme.
		Comparing of reconstructed images,
		we can clearly and accurately identify the number ``$7$" from the recovered images using the proposed robust $\beta$-VAE scheme.

		\begin{table*}[ht]
			\caption{Performance of the proposed semantic communication prototype on MNIST dataset}
			\centering
			\begin{tabular}{ | c | c | c | c | c|}
				\hline
				{ } & \tabincell{c}{Transmission\\ time (ms) } & \tabincell{c}{Compression\\ ratio } &PSNR& \tabincell{c}{Reconstructed\\ image } \\ \hline
				Original image& 18.86&1&100
				&\begin{minipage}[b]{0.1\columnwidth}
					\centering
					 \raisebox{-.5\height}{\includegraphics[width=\linewidth]{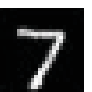}}
				\end{minipage}
				\\ \hline
				
				JPEG& 10.88& 1.81&17.49&\begin{minipage}[b]{0.1\columnwidth}
					\centering
					\raisebox{-.5\height}{\includegraphics[width=\linewidth]{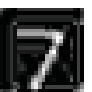}}
				\end{minipage}
				\\ \hline
				$\beta$-VAE 	& 0.30	& 78.4&15.79&\begin{minipage}[b]{0.1\columnwidth}
					\centering
					\raisebox{-.5\height}{\includegraphics[width=\linewidth]{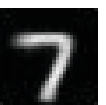}}
				\end{minipage}
				\\ \hline
				Robust $\beta$-VAE & 0.30	&78.4&16.07&\begin{minipage}[b]{0.1\columnwidth}
					\centering
					\raisebox{-.5\height}{\includegraphics[width=\linewidth]{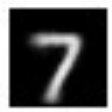}}
				\end{minipage}
				\\ \hline
				
			\end{tabular}\label{prototype_MNIST}
		\end{table*}
		\begin{table*}[ht]
	\caption{Performance of the proposed semantic communication prototype on CelebA dataset}
	\centering
	\begin{tabular}{ | c | c | c | c | c|}
		\hline
		{ } & \tabincell{c}{Transmission\\  time (ms) }  & \tabincell{c}{Compression\\ ratio }  &PSNR& \tabincell{c}{Reconstructed\\ image }  \\ \hline
		Original image& 28.38&1&100
		&\begin{minipage}[b]{0.1\columnwidth}
			\centering
			\raisebox{-.5\height}{\includegraphics[width=\linewidth]{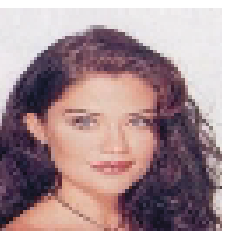}}
		\end{minipage}
		\\ \hline
		
		JPEG& 9.53& 4.49&30.17&\begin{minipage}[b]{0.1\columnwidth}
			\centering
			\raisebox{-.5\height}{\includegraphics[width=\linewidth]{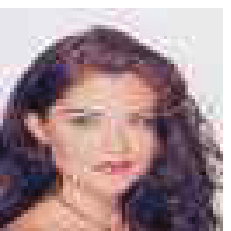}}
		\end{minipage}
		\\ \hline
		$\beta$-VAE 	& 0.18	& 384&17.66&\begin{minipage}[b]{0.1\columnwidth}
			\centering
			\raisebox{-.5\height}{\includegraphics[width=\linewidth]{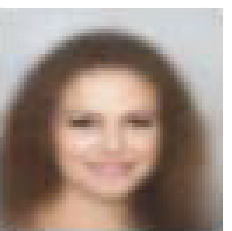}}
		\end{minipage}
		\\ \hline
		Robust $\beta$-VAE & 0.18	&384&19.73&\begin{minipage}[b]{0.1\columnwidth}
			\centering
			\raisebox{-.5\height}{\includegraphics[width=\linewidth]{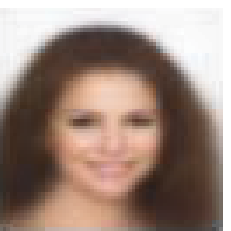}}
		\end{minipage}
		\\ \hline
		
	\end{tabular}\label{prototype_CelebA}
\end{table*}

	Table \ref{prototype_CelebA} compares a compression ratio, transmission time, PSNR and reconstructed image of the original image transmission scheme, JPEG compression scheme, $\beta$-VAE scheme, and robust $\beta$-VAE scheme over our proposed semantic communication prototype on CelebA dataset  with high SNR. Similar to Table \ref{prototype_MNIST},   the compression ratio of the $\beta$-VAE scheme and robust $\beta$-VAE scheme is $384$ which is significantly higher than those of the JPEG compression scheme (4.49) and original image transmission scheme. Thus, the transmission time of the $\beta$-VAE scheme and robust $\beta$-VAE scheme is about $0.18$ms, which is significantly lower than those of the JPEG compression scheme (9.53ms) and original image transmission scheme (28.38ms). Therefore, the proposed semantic communication system can significantly reduce the transmission load and time.
		Moreover, the PSNR of the robust $\beta$-VAE scheme is close to that of the JPEG compression scheme, and is higher than that of the $\beta$-VAE scheme.
Note that,
although the effect of the reconstructed image  of proposed robust $\beta$-VAE scheme is a bit blurry, the three main semantic features of the original   image: female, white skin color and long hair,  are all accurately transmitted, which verifies the validity and accuracy of the proposed task-oriented semantic communication scheme.


		\section{Conclusions}
		In this paper, we propose an explainable and easy-to-implement semantic communication framework that is compatible with conventional communication systems.
		In this new framework, the semantic encoder can extract feature vectors, disentangle
		the semantic information, and improve robustness against semantic information ambiguity.
		To further reduce the communication cost, we apply feature selection to choose only
		task-related semantic information to transmit.
		Then, we present two information theoretic metrics, namely, the rate-distortion-perception function and
		semantic channel capacity to characterize the semantic information compression and transmission,
		respectively. To quantify the semantic information transmission with the additive quantization noise
		and physical channel noise, we further derive upper and lower bounds on the semantic channel capacity.
		Then, we propose a feasible design of the explainable semantic communication system,
		which includes a robust $\beta$-VAE lightweight unsupervised learning network.
		Finally, we develop a wireless mobile semantic communication proof-of-concept prototype to implement the semantic communication design.
		Our experiments demonstrate that the proposed semantic communication system significantly
		outperforms
		the state-of-the-art methods, and shows robustness against various noise levels on two benchmark datasets.
		This work attempts to provide frameworks and theoretic metrics to explain and analyze the black-box semantic
		communications problem, and to provide guidelines on implementing the semantic communication in practical communication systems.

		\section{Appendices}
		
		\begin{appendices}

			\section{Proof of Lemma \ref{BAoptimalDistr}}
			{
				We first derive the optimal conditional distribution $q(\widehat x|x)$ in \eqref{R_D_P} for a given output distribution $r(x)$.
				The mutual information $I ( {X;\widehat X} ) = \sum\limits_x {\sum\limits_{\widehat x} {p\left( x \right)q( {\widehat x|x} )\log \frac{{q( {\widehat x|x} )}}{{r( {\widehat x}) }}} } $ is convex in ${q( {\widehat x|x} )}$ for fixed ${p\left( x \right)}$, and the KL divergence ${d_{KL}}\left( {p\left( x \right),r( {\widehat x}) } \right) = \sum\limits_x {p\left( x \right)\log \frac{{p\left( x \right)}}{{r( { x}) }}} $ is also convex in ${q( {\widehat x|x} )}$ for fixed ${p\left( x \right)}$.
				Thus, problem \eqref{R_D_P} is convex in ${q( {\widehat x|x} )}$.
				Then, the Lagrangian
				function of problem \eqref{R_D_P} is given by
				\begin{IEEEeqnarray}{rCl}
					L\left( {q( {\widehat x|x} )} \right) = &&~I\left( {X;\widehat X} \right) + \alpha {\sum\limits_x {\sum\limits_{\widehat x} {p(x)q(\widehat x|x)\left( {x - \widehat x} \right)} } ^2} \nonumber\\&&+ \mu \sum\limits_x {p\left( x \right)\log \frac{{p\left( x \right)}}{{r( {\widehat x}) }}} + \sum\limits_x {\gamma \left( x \right)\sum\limits_{\widehat x} {q( {\widehat x|x} )} }, \label{R_D_P_L}
				\end{IEEEeqnarray}
				where $\alpha \ge 0 $, $\mu \ge 0 $ and ${\gamma \left( x \right)}\ge 0 $ are Lagrange multipliers attached with constraints \eqref{R_D_P_b1}, \eqref{R_D_P_c1} and \eqref{R_D_P_d1}, respectively. For given $r(\widehat x)$, the derivative of \eqref{R_D_P_L} with respect to ${q( {\widehat x|x} )}$ is given as
				\begin{IEEEeqnarray}{rCl}
					\frac{\partial L\big( q( {\widehat x|x} ) \big)}{\partial q( {\widehat x|x} )} &&=~ p ( x )\big( \log \frac{q( {\widehat x|x} )}{r( {\widehat x}) } + \alpha {{\left( {x - \widehat x} \right)}^2} - \mu \frac{{p\left( x \right)}}{{r( {\widehat x}) }} + \frac{{\gamma \left( x \right)}}{{p\left( x \right)}} \big).
				\end{IEEEeqnarray}
				
				Let $\frac{{\partial L\left( {q( {\widehat x|x} )} \right)}}{{\partial q( {\widehat x|x} )}} = 0$, then we obtain the optimal $q( {\widehat x|x} )$ as
				\begin{subequations}
					\begin{align}
						q^* ( {\widehat x|x} ) &= r( {\widehat x}) \exp \left( {\mu \frac{{p\left( x \right)}}{{r( {\widehat x}) }} - \alpha {{\left( {x - \widehat x} \right)}^2} - \frac{{\gamma \left( x \right)}}{{p\left( x \right)}}} \right)\\
						&= \frac{{r( {\widehat x}) }}{{\widetilde \gamma \left( x \right)}}\exp \left( {\mu \frac{{p\left( x \right)}}{{r( {\widehat x}) }} - \alpha {{\left( {x - \widehat x} \right)}^2}} \right), \label{qXdist}
					\end{align}
				\end{subequations}
				where $\widetilde \gamma \left( x \right) \buildrel \Delta \over = \exp \left( {\frac{{\gamma \left( x \right)}}{{p\left( x \right)}}} \right)$.
				
				Since $\sum\limits_{\widehat x} {q( {\widehat x|x} )} = 1$,
				we have \begin{align}\sum\limits_{\widehat x} {\frac{{r( {\widehat x}) }}{{\widetilde \gamma \left( x \right)}}\exp \left( {\mu \frac{{p\left( x \right)}}{{r( {\widehat x}) }} - \alpha {{\left( {x - \widehat x} \right)}^2}} \right)} = 1.
				\end{align}
				Furthermore, we obtain \begin{IEEEeqnarray}{rCl}\label{gmmatildX}
					\widetilde \gamma \left( x \right) = \sum\limits_{\widehat x} {r( {\widehat x}) \exp \left( {\mu \frac{{p\left( x \right)}}{{r( {\widehat x}) }} - \alpha {{\left( {x - \widehat x} \right)}^2}} \right)}.
				\end{IEEEeqnarray}
				Substituting \eqref{gmmatildX} into \eqref{qXdist}, we obtain the optimal $q^*( {\widehat x|x} )$ as given in Lemma \ref{BAoptimalDistr}.
				
				From \cite{Cover_Book} , we find that given a fixed conditional distribution $q({\widehat x|x})$, the optimal output distribution $r(\widehat x)$ is $r^*(x)\triangleq \sum_{x}p(x)q({\widehat x|x})$. We rewrite the proof below.
				\begin{IEEEeqnarray}{rCl}
					I(X;Z) =&&~\sum_{x,\widehat x} p(x)q(\widehat x|x)\log \frac{p(x)q(\widehat x|x)}{p(x)r(\widehat x)}
					\\&&- \sum_{x,\widehat x} p(x)q(\widehat x|x)\log \frac{p(x)q(\widehat x|x)}{p(x)r^*(\widehat x)}
					\\ && {=} \sum_{\widehat x} r^*({x})\log \frac{r^*(\widehat x)}{r(\widehat x)} \geq 0,
				\end{IEEEeqnarray}
				where the last inequality holds because of the non-negative property of KL divergence.
			}

			\end{appendices}
			\bibliographystyle{IEEE-unsorted}
			\bibliographystyle{IEEEtran}

			\bibliography{refs0611} 

\begin{thebibliography}{10}

\bibitem{Lueth_2020}
L.~Knud,
\newblock ``State of the {IoT} 2020: 12 billion {IoT} connections, surpassing
  non-{IoT} for the first time,''
  \url{https://iot-analytics.com/state-of-the-iot-2020-12-billion-iot-connections-surpassing-non-iot
  /}, 2020.

\bibitem{Antoniou_2021}
J.~Antoniou,
\newblock ``Quality of experience and emerging technologies: Considering
  features of {5G}, {IoT}, cloud and {AI},''
\newblock in {\em Quality of Experience and Learning in Information Systems},
  pp. 1--8. Springer, 2021.

\bibitem{Saad_INW_2020}
W.~Saad, M.~Bennis, and M.~Chen,
\newblock ``A vision of {6G} wireless systems: Applications, trends,
  technologies, and open research problems,''
\newblock {\em IEEE Netw.}, vol.~34, no.~3, pp.~134--142, Oct. 2020.

\bibitem{Strinati_TVT_2019}
E.~Calvanese~Strinati, S.~Barbarossa, J.~L. Gonzalez-Jimenez, D.~Ktenas,
  N.~Cassiau, L.~Maret, and C.~Dehos,
\newblock ``{6G}: The next frontier: From holographic messaging to artificial
  intelligence using subterahertz and visible light communication,''
\newblock {\em IEEE Veh. Technol. Mag.}, vol.~14, no.~3, pp.~42--50, Oct. 2019.

\bibitem{Bao_ICST_2022}
B.~Mao, F.~Tang, Y.~Kawamoto, and N.~Kato,
\newblock ``{AI} models for green communications towards {6G},''
\newblock {\em IEEE Commun. Surveys Tuts.}, vol.~24, no.~1, pp.~210--247, Nov.
  2022.

\bibitem{Niu_arXiv_2022}
K.~Niu, J.~Dai, S.~Yao, S.~Wang, Z.~Si, X.~Qin, and P.~Zhang,
\newblock ``Towards semantic communications: A paradigm shift,''
\newblock {\em arXiv preprint arXiv:2203.06692}, 2022.

\bibitem{Zhang_Engineering_2022}
P.~Zhang, W.~Xu, H.~Gao, K.~Niu, X.~Xu, X.~Qin, C.~Yuan, Z.~Qin, H.~Zhao,
  J.~Wei, et~al.,
\newblock ``Toward wisdom-evolutionary and primitive-concise {6G}: A new
  paradigm of semantic communication networks,''
\newblock {\em Engineering}, 2022.

\bibitem{Kountouris_CM_2021}
M.~Kountouris and N.~Pappas,
\newblock ``Semantics-empowered communication for networked intelligent
  systems,''
\newblock {\em IEEE Commun. Mag.}, vol.~59, no.~6, pp.~96--102, Jan. 2021.

\bibitem{Sana_CCNC_2022}
M.~Sana and E.~Calvanese~Strinati,
\newblock ``Learning semantics: An opportunity for effective {6G}
  communications,''
\newblock {\em arXiv preprint arXiv:2202.11958}, 2021.

\bibitem{Shi_CM_2021}
Y.~L. G.~Shi~Y.~Xiao. and X.~Xie,
\newblock ``From semantic communication to semantic-aware networking: Model,
  architecture, and open problems,''
\newblock {\em IEEE Commun. Mag.}, vol.~59, no.~8, pp.~44--50, Aug. 2021.

\bibitem{Luo_WC_2022}
X.~Luo, H.-H. Chen, and Q.~Guo,
\newblock ``Semantic communications: Overview, open issues, and future research
  directions,''
\newblock {\em IEEE Wirel. Commun.}, pp. 1--10, Jan. 2022.

\bibitem{Bao_INSW_2011}
J.~Bao, P.~Basu, M.~Dean, C.~Partridge, A.~Swami, W.~Leland, and J.~A. Hendler,
\newblock ``Towards a theory of semantic communication,''
\newblock in {\em Proc. IEEE Netw. Sci. Workshop}, pp. 110--117, Jun. 2011.

\bibitem{Yener_TCCN_2018}
A.~Y. B.~G{\"u}ler and A.~Swami,
\newblock ``The semantic communication game,''
\newblock {\em IEEE Trans. Cogn. Commun. Netw.}, vol.~4, no.~4, pp.~787--802,
  Dec. 2018.

\bibitem{Weaver_1949}
W.~Weaver,
\newblock ``Recent contributions to the mathematical theory of communication,''
\newblock {\em ETC: a review of general semantics}, pp. 261--281, Sep. 1949.

\bibitem{Shannon_1948}
C.~E. Shannon,
\newblock ``A mathematical theory of communication,''
\newblock {\em Bell Syst. Tech. J.}, vol.~27, no.~3, pp.~623--656, Jul. 1948.

\bibitem{Farsad_ICASSP_2018}
N.~Farsad, M.~Rao, and A.~Goldsmith,
\newblock ``Deep learning for joint source-channel coding of text,''
\newblock in {\em Proc.(ICASSP)}, pp. 2326--2330, Apr. 2018.

\bibitem{Xie_TSP_2021}
H.~Xie, Z.~Qin, L.~Geoffrey~Ye., and B.-H. Juang,
\newblock ``Deep learning enabled semantic communication systems,''
\newblock {\em IEEE Trans. Signal Process.}, vol.~69, pp.~2663--2675, Apr.
  2021.

\bibitem{Hu_arXiv_2022}
Q.~Hu, G.~Zhang, Z.~Qin, Y.~Cai, and G.~Yu,
\newblock ``Robust semantic communications against semantic noise,''
\newblock {\em arXiv preprint arXiv:2202.03338}, Feb. 2022.

\bibitem{Bourtsoulatze_TCCN_2019}
E.~Bourtsoulatze, D.~B. Kurka, and D.~G{\"u}nd{\"u}z,
\newblock ``Deep joint source-channel coding for wireless image transmission,''
\newblock {\em IEEE Trans. Cognit.Commun. Netw.}, vol.~5, no.~3, pp.~567--579,
  May. 2019.

\bibitem{Kurika_JSAC_2020}
D.~B. Kurka and D.~G{\"u}nd{\"u}z,
\newblock ``Deepjscc-f: Deep joint source-channel coding of images with
  feedback,''
\newblock {\em IEEE J. Sel. Areas Inf. Theory}, vol.~1, no.~1, pp.~178--193,
  Apr. 2020.

\bibitem{Yang_TCCN_2022}
M.~Yang, C.~Bian, and H.-S. Kim,
\newblock ``{OFDM}-guided deep joint source channel coding for wireless
  multipath fading channels,''
\newblock {\em IEEE Trans. Cognit. Commun. Netw.}, Feb. 2022.

\bibitem{Choi_PMLR_2019}
K.~Choi, K.~Tatwawadi, A.~Grover, T.~Weissman, and S.~Ermon,
\newblock ``Neural joint source-channel coding,''
\newblock in {\em Proc. Int. Conf. Mach. Learn.(ICML)}. PMLR, pp. 1182--1192,
  Jun. 2019.

\bibitem{Xu_TCSVT_2021}
J.~Xu, B.~Ai, W.~Chen, A.~Yang, P.~Sun, and M.~Rodrigues,
\newblock ``Wireless image transmission using deep source channel coding with
  attention modules,''
\newblock {\em IEEE Trans. Circuits Syst. Video Technol.}, May. 2021.

\bibitem{Tong_GLOBECOM_2021}
H.~Tong, Z.~Yang, S.~Wang, Y.~Hu, W.~Saad, and C.~Yin,
\newblock ``Federated learning based audio semantic communication over wireless
  networks,''
\newblock in {\em Proc. IEEE Global Commun. Conf. (GLOBECOM)}, pp. 1--6, Feb.
  2021.

\bibitem{Weng_JSAC_2021}
Z.~Weng and Z.~Qin,
\newblock ``Semantic communication systems for speech transmission,''
\newblock {\em IEEE J. Sel. Areas Commun.}, vol.~39, no.~8, pp.~2434--2444,
  Aug. 2021.

\bibitem{Xie_WCL_2021}
H.~Xie, Z.~Qin, and G.~Y. Li,
\newblock ``Task-oriented multi-user semantic communications for {VQA},''
\newblock {\em IEEE Wirel. Commun. Lett.}, vol.~11, no.~3, pp.~553--557, Dec.
  2021.

\bibitem{Zhou_arxiv_2022}
Q.~Zhou, R.~Li, Z.~Zhao, Y.~Xiao, and H.~Zhang,
\newblock ``Adaptive bit rate control in semantic communication with
  incremental knowledge-based {HARQ},''
\newblock {\em arXiv preprint arXiv:2203.06634}, 2022.

\bibitem{Jiang_arXiv_2021}
P.~Jiang, C.-K. Wen, S.~Jin, and G.~Y. Li,
\newblock ``Deep source-channel coding for sentence semantic transmission with
  {HARQ},''
\newblock {\em arXiv preprint arXiv:2106.03009}, 2021.

\bibitem{Zhou_WCL_2021}
Z.~Z. C.~P. Q.~Zhou~R.~Li. and H.~Zhang,
\newblock ``Semantic communication with adaptive universal transformer,''
\newblock {\em IEEE Wirel. Commun. Lett.}, vol.~11, no.~3, pp.~453--457, Dec.
  2021.

\bibitem{Xie_JSAC_2021}
H.~Xie and Z.~Qin,
\newblock ``A lite distributed semantic communication system for internet of
  things,''
\newblock {\em IEEE J. Sel. Areas Commun.}, vol.~39, no.~1, pp.~142--153, Jan.
  2021.

\bibitem{Lu_arxiv_2021}
K.~Lu, R.~Li, X.~Chen, Z.~Zhao, and H.~Zhang,
\newblock ``Reinforcement learning-powered semantic communication via semantic
  similarity,''
\newblock {\em arXiv preprint arXiv:2108.12121}, 2021.

\bibitem{Yang_arXiv_2021}
M.~Yang and H.-S. Kim,
\newblock ``Deep joint source-channel coding for wireless image transmission
  with adaptive rate control,''
\newblock {\em arXiv preprint arXiv:2110.04456}, 2021.

\bibitem{Ding_ICASSP_2021}
M.~Ding, J.~Li, M.~Ma, and X.~Fan,
\newblock ``{SNR}-adaptive deep joint source-channel coding for wireless image
  transmission,''
\newblock in {\em Proc. IEEE Int. Conf. Acoust., Speech, Signal
  Process.(ICASSP)}, pp. 1555--1559, May. 2021.

\bibitem{Shao_JSAC_2022}
Y.~M. J.~Shao and J.~Zhang,
\newblock ``Learning task-oriented communication for edge inference: An
  information bottleneck approach,''
\newblock {\em IEEE J. Sel. Areas Commun.}, vol.~40, no.~1, pp.~197--211, Jan.
  2022.

\bibitem{Blau_ICML_2019}
Y.~Blau and T.~Michaeli,
\newblock ``Rethinking lossy compression: The rate-distortion-perception
  tradeoff,''
\newblock in {\em International Conference on Machine Learning}. PMLR, pp.
  675--685, 2019.

\bibitem{Zhang_NIPS_2021}
G.~Zhang, J.~Qian, J.~Chen, and A.~Khisti,
\newblock ``Universal rate-{Distortion}-{Perception} representations for lossy
  compression,''
\newblock {\em Advances in Neural Information Processing Systems}, vol.~34,
  2021.

\bibitem{Cover_Book}
T.~M. Cover and J.~A. Thomas,
\newblock {\em Elements of information theory, 2nd ed.},
\newblock New York, NY, USA: Wiley, 2006.

\bibitem{Ihara}
S.~Ihara,
\newblock ``On the capacity of channels with additive {non-Gaussian} noise,''
\newblock {\em Inform. Contr.}, vol.~37, no.~1, pp.~34--39, Sep. 1978.

\bibitem{Higgins_ICLR_2017}
I.~Higgins, L.~Matthey, A.~Pal, C.~Burgess, X.~Glorot, M.~Botvinick,
  S.~Mohamed, and A.~Lerchner,
\newblock ``beta-vae: Learning basic visual concepts with a constrained
  variational framework,''
\newblock {\em Proc. ICLR}, pp. 1--12, 2017.

\bibitem{Kingma_arXiv}
D.~P. Kingma and M.~Welling,
\newblock ``Auto-encoding variational bayes,''
\newblock {\em arXiv preprint arXiv:1312.6114}, 2013.

\bibitem{Eguchi_Entropy_2010}
S.~Eguchi and S.~Kato,
\newblock ``Entropy and divergence associated with power function and the
  statistical application,''
\newblock {\em Entropy}, vol.~2, pp.~262?274, Dec. 2010.

\bibitem{Futami_arxiv_2017}
F.~Futami, I.~Sato, and M.~Sugiyama,
\newblock ``Variational inference based on robust divergences,''
\newblock {\em arXiv preprint arXiv:1710.06595}, 2017.

\end{thebibliography}
			
		\end{document}